\documentclass[aps,10pt,prb,amsmath,superscriptaddress,amssymb,longbibliography,twocolumn]{revtex4-2}

\usepackage[bottom]{footmisc}
\usepackage{amsmath} 
\usepackage{graphicx}
\usepackage{booktabs}
\usepackage{dcolumn}
\usepackage{bm}
\usepackage[	
citecolor=blue,
colorlinks, 
linkcolor=red,
urlcolor=magenta,
]{hyperref}
\usepackage{braket}
\usepackage{multirow}
\usepackage{appendix}
\usepackage{wasysym}
\usepackage{slashed}
\usepackage[table, svgnames, dvipsnames]{xcolor}
\usepackage{amsfonts}
\usepackage{soul}

\usepackage{subfigure}

\newcommand{\mcl}[1]{\mathcal{#1}}

\usepackage{tikz}

\begin{document}

\title{Phases and phase transitions in a dimerized spin-\texorpdfstring{$\mathbf{\frac{1}{2}}$}{} XXZ chain}

\author{Harsh Nigam}
\email{harsh.nigam@icts.res.in}
\affiliation{International Centre for Theoretical Sciences, Tata Institute of Fundamental Research, Bangalore 560089, India}
\author{Ashirbad Padhan}
\email{padhanashirbad@gmail.com}
\affiliation{School of Physical Sciences, National Institute of Science Education and Research, Jatni 752050, India}
\affiliation{Homi Bhabha National Institute, Training School Complex, Anushaktinagar, Mumbai 400094, India}

\author{Diptiman Sen}
\email{diptiman@iisc.ac.in}
\affiliation{Centre for High Energy Physics, Indian Institute of Science, Bengaluru 560012, India}

\author{Tapan Mishra}
\email{mishratapan@niser.ac.in}
\affiliation{School of Physical Sciences, National Institute of Science Education and Research, Jatni 752050, India}
\affiliation{Homi Bhabha National Institute, Training School Complex, Anushaktinagar, Mumbai 400094, India}

\author{Subhro Bhattacharjee}
\email{subhro@icts.res.in}
\affiliation{International Centre for Theoretical Sciences, Tata Institute of Fundamental Research, Bangalore 560089, India}

\begin{abstract}
We revisit the phase diagram of the dimerized XXZ spin-$\frac{1}{2}$ chain with
nearest-neighbor couplings which was studied numerically in \href{https://link.aps.org/doi/10.1103/PhysRevB.106.L201106}{Phys. Rev. B 106, L201106 (2022)}.  The model has isotropic $XY$ couplings which have a uniform value and $ZZ$ couplings  which have a dimerized form, with strengths $J_a$ and $J_b$ on alternate bonds. We find a rich phase diagram in the region of positive $J_a, ~J_b$. We provide a detailed understanding of the different phases and associated quantum phase  transitions using a combination of mean-field theory, low-energy effective Hamiltonians, renormalization group calculations employing the technique of  bosonization, and numerical calculations using the density-matrix renormalization group (DMRG) method. The phase diagram consists of two Ising paramagnetic phases  called IPM$_0$ and IPM$_\pi$, and a phase with Ising Neel order called IN; all these phases are gapped. The phases IPM$_0$ and IPM$_\pi$ are separated by a gapless phase transition line given by $0 \le J_a = J_b \le 1$ which is described by a conformal field theory with central charge $c=1$. There are two gapless phase transition lines separating IPM$_0$ from IN and IPM$_\pi$ from IN; these are described by conformal field theories with $c=\frac{1}{2}$ corresponding to quantum Ising transitions. The $c=1$ line bifurcates into the two $c=\frac{1}{2}$ lines at the point $J_a = J_b = 1$; the shape of the bifurcation is found analytically using RG calculations. A symmetry analysis shows that IPM$_0$ is a topologically trivial phase while IPM$_\pi$ is a  time-reversal symmetry-protected topological phase (SPT) with spin-$\frac{1}{2}$ states at the two ends of an open system. A gap opens in the low-energy spectrum whenever one moves away from one of the phase transition lines; the scaling of the gap with the distance from a transition line is found analytically using the RG method. The numerical results obtained by the DMRG method are in good agreement with the analytical results. Further, we show that useful insights into the nature of the phase diagram of the above Hamiltonian, particularly the SPT as well as the phase transitions, is obtained via Kramers-Wannier duality to a deformed quantum Ashkin-Teller model. Finally we propose experimental platforms for testing our results.
\end{abstract}

\maketitle

\section{Introduction}

Quantum spin chains have attracted renewed attention recently in the context of classification of symmetry-protected topological (SPT) phases~\cite{PhysRevB.85.075125,chen2012symmetry,SPT_Wen,senthil2015symmetry}, various unconventional quantum phase transitions beyond the conventional Landau-Ginzburg paradigm~\cite{PhysRevB.99.075103,PhysRevB.99.205153,PhysRevB.99.165125,PhysRevB.103.195134} and very recently in the context of lattice examples of unconventional realization of symmetries including generalized non-invertible symmetries~\cite{seiberg2024non,seifnashri2024cluster,lu2024realizing,seiberg2024majorana,cordova2024particle,PhysRevB.108.214429}. In particular the 
spin-$\frac{1}{2}$ XXZ chains, in various versions, continue to serve as a minimal model to explore these ideas and systematically bridge the microscopic physics with the low-energy effective field theories for both the phases as well as the associated phase transitions. In a parallel exciting development, various one dimensional correlated  Hamiltonians and associated transitions have been realized in experiments in Rydberg atom arrays  where range of inter-particle interactions can be tuned by changing the Rydberg blocking radius~\cite{PhysRevA.98.023614,browaeys2020many,bernien2017probing,PhysRevB.69.075106,chepiga2021kibble,PhysRevResearch.4.043102}.

\begin{figure}
    \centering
\includegraphics[width=1\columnwidth]{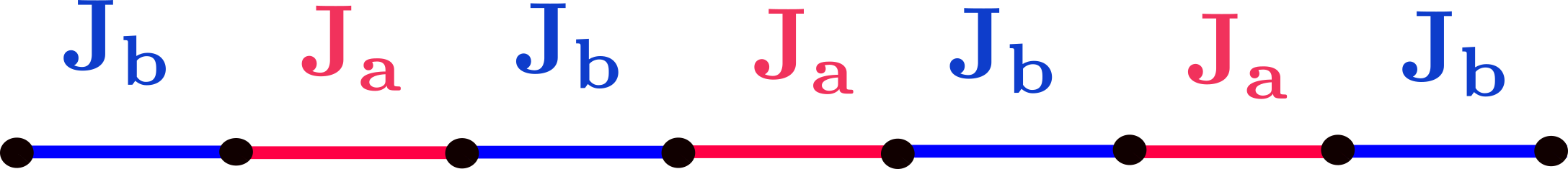}
    \caption{A schematic of spin chain of Hamiltonian in Eq.~\ref{eqn:eq2.1}. A spin-$\frac{1}{2}$ resides on sites labelled by index $i$ (black dots). The $ZZ$ exchanges are dimerized in the sense that they alternate between $J_a $ and $J_b$ on even (red) and odd (blue) bonds respectively.}
    \label{fig:fig3}
\end{figure}

\begin{figure}[htb]
\centering
\includegraphics[width=1\columnwidth]{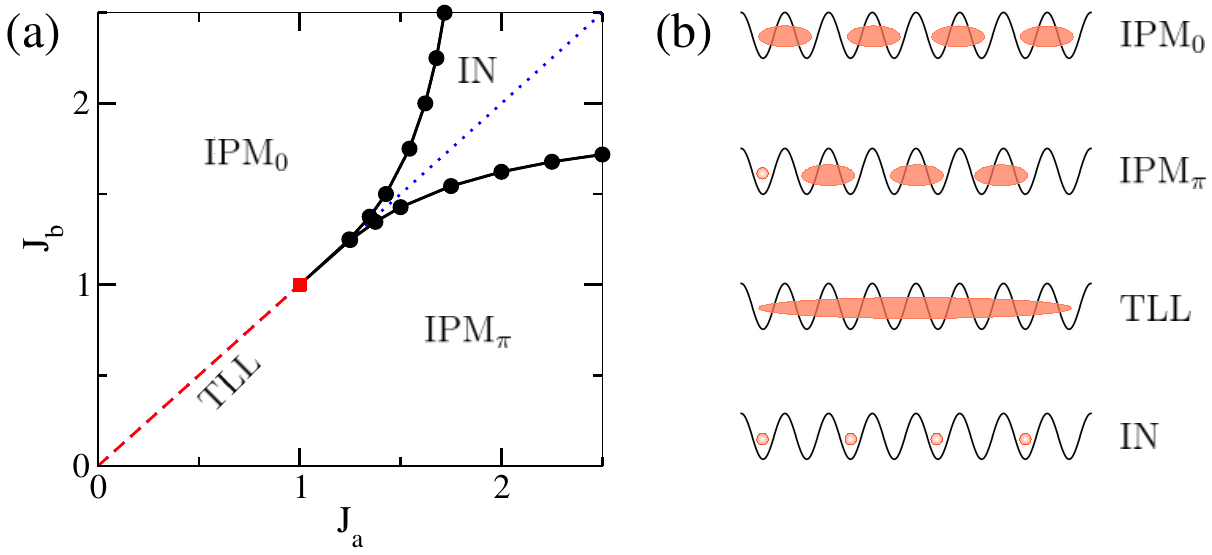}
\caption{(a) Ground state phase diagram of the model in Eq.~\ref{eq:ham_v1v2} in the $J_a-J_b$ plane for $J_{\perp}=1$, which is obtained using the DMRG method. The red dashed line represents the gapless $XY$-ordered phase. The red square denotes the well-known Heisenberg point for the transition from the $XY$-ordered phase to the Neel phase.  (b) Shows the  schematic for the distribution of spins in all the phases depicted in (a).}

\label{fig:pd}
\end{figure}

In this paper, following the recent numerical results of Ref.~\cite{PhysRevB.106.L201106}, we revisit the dimerized XXZ spin-$\frac{1}{2}$ chain and provide an understanding of numerically obtained phase diagram (Fig.~\ref{fig:pd}) consisting of interacting symmetry protected topological and trivial  as well as spontaneous symmetry broken phases including the intervening phase transitions. An application of the Kramers-Wannier (KW)~\cite{PhysRev.60.252} duality shows that this dimerized XXZ spin chain is related to modified quantum Ashkin-Teller (QAT)~\cite{PhysRev.64.178,PhysRevB.24.5229} and cluster Ising~\cite{PhysRevB.91.165129} models with non-manifest realization of symmetries (of the XXZ model)~\cite{PhysRevB.108.125135}.

The Hamiltonian for the dimerized XXZ spin-chain is given by~\cite{PhysRevB.106.L201106}
\begin{equation}\label{eqn:eq2.1}
H=-J_\perp \sum_i\left(S_i^xS_{i+1}^x+S_i^yS_{i+1}^y\right)+\sum_i J_{\parallel,i}\left(S_i^zS_{i+1}^z\right),
\end{equation}
where, as shown in Fig.~\ref{fig:fig3}, 
\begin{equation}\label{eqn:eq2.2}
J_{\parallel} =\begin{cases}
J_{a} \ \ \ \ \ \ \forall \ i\in \ \text{even}\\
J_{b} \ \ \ \ \ \ \forall \ i\in \ \text{odd}
\end{cases}
\end{equation}
and $S^\alpha_i ~(\alpha=x,y,z)$ are spin-$\frac{1}{2}$'s sitting on the sites of the chain (Fig.~\ref{fig:fig3}). Notably while the transverse coupling, $J_\perp(>0)$ has full translation symmetry of the chain, the longitudinal couplings, $J_\parallel (>0)$, alternate between $J_a $ and $J_b$ on even and odd bonds respectively. In this sense the above model is different from the dimerized XXZ models studied previously~\cite{affleck1989fields,PhysRevB.108.245135, PhysRevB.109.235143, juliàfarré2024quantizedthoulesspumpsprotected} where even the transverse interactions were dimerized leading to a SPT phase even in the pure $XX$ limit whence the Hamiltonian reduces to the well known Su--Schrieffer-Heeger (SSH) model~\cite{PhysRevLett.42.1698,tasaki2023rigorous}.  As discussed in Ref.~\cite{PhysRevB.106.L201106}, since the dimerization is in the ZZ interaction, the Hamiltonian above is not trivially connected to the dimerized XX Hamiltonian or equivalently SSH model by turning off the $ZZ$ interactions. Instead results in a gapless state and the ZZ interactions is responsible for dimerization and the associated SPT at intermediate values of $J_a$ and $J_b$. Thus the SPT phase and transition does not inherit from the non-interacting limit of the system. On the 
un-dimerized line,  $J_a=J_b$, the above Hamiltonian reduces to the regular XXZ 
spin-$\frac{1}{2}$ chain which has a transition between the gapless Tomonaga-Luttinger liquid (TLL)~\cite{haldane1981luttinger,Giamarchi_book,sachdev_2011} phase and an Ising Neel phase on monotonically increasing the longitudinal interactions. Thus the dimerized longitudinal interaction opens up a knob to understand the interplay of the SPT as well as spontaneous symmetry breaking.

Ref.~\cite{PhysRevB.106.L201106} employed density matrix renormalization group (DMRG) to numerically obtain the phase diagram of the spin Hamiltonian in Eq.~\ref{eqn:eq2.1} as a function of $J_a/J_b$ when $J_a,J_b<J_\perp$. (In numerical studies, the well known bosonic representation for the spins was used and this is summarized in Appendix~\ref{appen_spin-boson}). Extending the same DMRG calculations to more generic parameter values over the $J_a-J_b$ plane (we take $J_\perp=1$) gives the phase diagram in Fig.~\ref{fig:pd}. Along the symmetric line ($J_a=J_b=J_\parallel$), Eq.~\ref{eqn:eq2.1} reduces to the well-known XXZ model which exhibits a gapless power-law $XY$ ordered phase that possesses quasi-long-range order for $J_\parallel\leq 1$  denoted by the red line; this is described as a TLL~\cite{haldane1981luttinger,Giamarchi_book,sachdev_2011}. Above this, the TLL gives way to a gapped Ising Neel (IN) order as shown in Fig.~\ref{fig:pd}. The transition between the two phases occurs at the well-known Heisenberg point (red square) and is described by a compact boson CFT with central charge, $c=1$~\cite{PhysRevB.84.195108}. However, on moving away from the symmetric line \textit{i.e.} $J_a\neq J_b$, results in different scenarios at half filling which, also, have been discussed in Ref.~\cite{PhysRevB.106.L201106} and is shown in Fig.~\ref{fig:pd}. For smaller $J_\parallel$, two different gapped Ising paramagnetic (IPM) phases emerge by introducing a finite dimerization   (\textit{i.e.} $J_a\neq J_b$) on either sides of the TLL. These two IPM phases exhibit distinct (symmetry protected) topological character. For an open system of the type shown in Fig.~\ref{fig:fig3},  while the IPM phase for $J_a>J_b$ is found to be an SPT with gapless spin-$\frac{1}{2}$ edge modes similar to the Haldane phase (IPM$_\pi$), the one that appears for $J_a < J_b$ is trivial (IPM$_0$). As already discussed in Ref.~\cite{PhysRevB.106.L201106}, the IPM$_\pi$ is characterized by vanishing single-particle excitation gap in an open system due to dangling spin-$\frac{1}{2}$ edge states that is equivalently characterized by a finite string order parameter and finite Berry phase~\cite{PhysRevB.106.L201106}. While there is a direct transition between the two IPMs for smaller $J_a,J_b$, {for larger $J_a, J_b$, each of the IPMs separately give way to the IN phase via a continuous phase transition as depicted in Fig.~\ref{fig:pd}.} These IPM-IN transitions were found to belong to the Ising universality class in Ref.~\cite{PhysRevB.106.L201106} from the finite-size scaling of the density structure factor. Further, as expected {for open chains}, due to the edge modes in the IPM$_\pi$, the single-particle (in terms of fermions/bosons) gap was found to close at its transition with IN, while no-such {\it single-particle} gap closing was observed for the case of IN-IPM$_0$ transition. Thus according to this picture, the $c=1$ critical point (between the TLL-IN) branches out into two Ising transitions described by $c=\frac{1}{2}$ CFTs. 

\begin{figure}
\centering
\includegraphics[width=1\columnwidth]{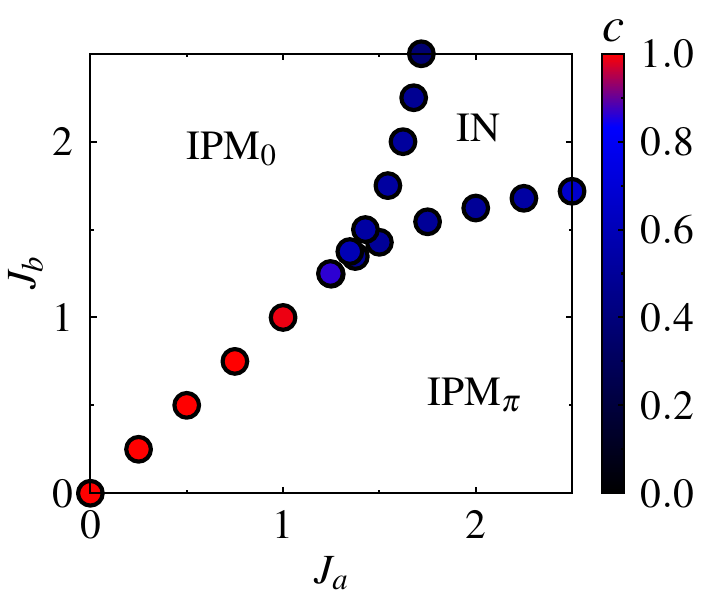}
\caption{Phase diagram in the $J_a-J_b$ plane obtained using the DMRG method for a system of size $L=1000$, with $J_\perp = 1$. Here, the color bar denotes the values of the central charge $c$ (scaled from $0$ to $1$). Details for the calculation of the central charge is given in Appendix~\ref{appen_centralcharge}.} 
\label{fig:pd_cc}
\end{figure}

Here, we revisit the phase diagram depicted in Fig.~\ref{fig:pd} via {duality, perturbative strong coupling expansions,  mean-field theories and renormalization group (RG) analysis of bosonized continuum low-energy field theories to uncover the rich interplay of symmetry and topology including the instability of the $c=1$ critical point to the two $c=\frac{1}{2}$ Ising critical lines. We show that the basic structure of the phase diagram as well as the phase transitions can be obtained by starting} with the anisotropic limit of $J_b$ (or $J_a)\rightarrow \infty$ whence the Hamiltonian (Eq.~\ref{eqn:eq2.1}) reduces to a transverse field Ising model at low energy in the leading order of $J_\perp/J_b, J_a/J_b$ (or $J_\perp/J_a, J_b/J_a)$. This leads (Fig.~\ref{fig:pd}) to IPM and IN phases as $J_a/J_\perp$ (or $J_b/J_\perp$) is increased with an intervening phase transition belonging to 2D Ising universality class and hence accounting for the $c=\frac{1}{2}$ central charge along this line that is now explicitly seen in numerical calculations (Fig.~\ref{fig:pd_cc}) along with the $c=1$ for the TLL as well as the phase transition along the symmetric line. Further, The IPM$_\pi$ obtained when $J_a/J_b>1$ is an SPT similar to the Haldane phase~\cite{haldane1983continuum} while IPM$_0$ obtained for $J_b/J_a>1$ is trivial. {The non-trivial character of IPM$_\pi$ is best revealed by employing the KW duality~\cite{PhysRev.60.252} that maps the above XXZ Hamiltonian (Eq.~\ref{eqn:eq2.1}) to a deformed quantum Ashkin-Teller model (QAT)~\cite{PhysRev.64.178,PhysRevB.24.5229} and the cluster Ising model~\cite{PhysRevB.91.165129} as shown in Appendix~\ref{app:XXZ_to_QAT}. Indeed, we show that the dual Hamiltonians admit a decorated domain-wall description~\cite{chen2014symmetry} for the SPT.

A complementary insight into the phase diagram is obtained via (spinless) fermionic representation of the spins at the mean-field (MF) level. Indeed the IPM$_\pi$ (IPM$_0$), at this MF level, is simply the topological (trivial) band insulator of the SSH model for the resultant fermions and the IN is a CDW insulator. The DMRG calculations show that all the phases are gapped in the bulk and the associated quantum phase transitions are continuous. In particular, the transition between the IPM phases and the IN phase for generic anisotropy belong to (1+1)D Ising universality class characterized by chiral CFTs with central charge $c=\frac{1}{2}$ (see Fig.~\ref{fig:pd_cc}). The two distinct Ising transitions from the IPM$_0$ and IPM$_\pi$ into the IN phase merge to a multi-critical point along the $J_a=J_b$ line that describes the regular transition between the TLL and the IN phase via a $c=1$ CFT for the regular XXZ model. Systematic derivation of the low-energy bosonized Hamiltonian (Eq.~\ref{eqn:total_bose_Ht}) leads to the {\it double-frequency sine-Gordon} model~\cite{DELFINO1998675,Fei_2003,fabrizio2000critical} that faithfully captures the above low-energy phenomenology-- the phases and the phase transitions.

The rest of the paper is organized as follows. We start with the anisotropic limits of $J_b$ (or $J_a)\rightarrow \infty$ to describe the two IPM phases and discuss their nature in Sec.~\ref{sec_IPM} using a degenerate perturbation theory. A complementary insight into the phase diagram is obtained by mapping the spins to spinless fermions through the Jordan-Wigner transformation. The resultant mean-field theory, which reproduces all the phases faithfully, is obtained in Sec.~\ref{sec_mft}. The DMRG results for the various quantum phase transitions are discussed in Sec.~\ref{sec_dmrg} while the critical theories and the phase diagram obtained via bosonization is discussed in Sec.~\ref{sec_bosonize}. We summarize our results in Sec.~\ref{sec_summary} while technical details are relegated to various appendices. {Not surprisingly, the same conclusion about the nature of the 
low-energy field theory can be obtained starting with the dual QAT~\cite{DELFINO1998675,fabrizio2000critical}.} Technical details are given in different appendices.


\section{Ising paramagnetic phases}
\label{sec_IPM}

Central to the understanding of the numerical phase diagram are the two IPM phases including the non-trivial nature of IPM$_\pi$ compared to IPM$_0$. This is best achieved by considering the two extremely dimerized limits-- $J_a\gg J_b$ and $J_a\ll J_b$. However, before we delve into these limits, it is useful to start with the symmetries of the system. These are
\begin{itemize}
    \item Time reversal, $\mathcal{Z}_2^T$, implemented on each spin by $\mathcal{T}=-i2 S^y \mathcal{K}$ where $\mathcal{K}$ is complex conjugation operator,
    \item Rotation about $S^z$, $\mathrm{U}_{z}(1)$, generated by $\mathcal{D}_z (\chi)=\exp(-i \mathcal{S}_z\chi)$ where $\mathcal{S}_z=\sum_i S_i^z$ is the total $z$-component of the spins and $\chi\in [0,2\pi]$,
    \item $\pi$-rotation about x-axis, $Z_2^x$, generated by $R_\pi^x=e^{-i\pi S_x}$. Note that  $[\mathcal{D}_z(\chi),R_\pi^x]\neq 0$, \textit{i.e.} the two symmetries do not commute and hence the symmetry group generated by these two is $U_z(1)\rtimes Z^x_2$,
    \item Reflection about the centre of $J_b$ bond, $\mathcal{R}_{J_b}$, 
    \item Lattice translation symmetry, $T_{2a}$, such that $x\rightarrow x+2a$, where $a$ is lattice parameter (see Fig.~\ref{fig:fig3}),
\end{itemize}
with the last two being spatial symmetries. The action of these symmetries on the spins is given in Table~\ref{tab:symm_tab_S} along with the corresponding action on the fermions (given by Eq.~\ref{eqn:eq5}).

    \begin{table*}[t]
        \centering
        
\begin{tabular}{p{0.15\textwidth} p{0.12\textwidth} p{0.14\textwidth} p{0.12\textwidth} p{0.12\textwidth} p{0.102\textwidth}}
\hline \hline \addlinespace
 Symmetry & $\displaystyle \mathcal{Z}_2^T$ & $\displaystyle U_{z}( 1)$ & $\displaystyle 
 R_{\pi}^{x}$ & $\mathcal{R}_{J_b}$  & $T_{2a}$\\
\hline
\addlinespace
 $\displaystyle S_{i}^{x}$ & $\displaystyle -S_{i}^{x}$ & $\displaystyle e^{i \mathcal{S}_z\chi} S_{i}^{x} e^{-i \mathcal{S}_z\chi}$ & $\displaystyle +S_{i}^{x}$ & $\displaystyle S_{L-i+1}^{x}$ & $\displaystyle S_{i+2}^{x}$ \\

 $\displaystyle S_{i}^{y}$ & $\displaystyle -S_{i}^{y}$ & $\displaystyle e^{i \mathcal{S}_z\chi} S_{i}^{y} e^{-i \mathcal{S}_z\chi}$ & $-\displaystyle S_{i}^{y}$ & $\displaystyle S_{L-i+1}^{y}$ & $\displaystyle S_{i+2}^{y}$ \\

 $\displaystyle S_{i}^{z}$ & $\displaystyle -S_{i}^{z}$ & $\displaystyle S_{i}^{z}$ & $-\displaystyle S_{i}^{z}$ & $\displaystyle S_{L-i+1}^{z}$ & $\displaystyle S_{i+2}^{z}$ \\ \addlinespace
$\hat{c}_{i}$ & $(-1)^i \hat{c}_{i}^\dag$ & $e^{-i\chi}\hat{c}_{i}$ & $(-1)^{i-1} \hat{c}_{i}^\dag$ & $\hat{c}_{L -i +1}$ & $\hat{c}_{i+2}$ \\ \addlinespace
\hline \addlinespace
 $\displaystyle \sigma _{I}^{x}$ & $\displaystyle \sigma _{I}^{x}$ & $\displaystyle \sigma _{I}^{x}$ & $\displaystyle \sigma _{I}^{x}$ & $\displaystyle \sigma _{L/2 +1-I}^{x}$ & $\displaystyle \sigma _{I+1}^{x}$ \\

 $\displaystyle \sigma _{I}^{y}$ & $\displaystyle \sigma _{I}^{y}$ & $\displaystyle \sigma _{I}^{y}$ & -$\displaystyle \sigma _{I}^{y}$ & -$\displaystyle \sigma _{L/2 +1-I}^{y}$ & $\displaystyle \sigma _{I+1}^{y}$ \\

 $\displaystyle \sigma _{I}^{z}$ & -$\displaystyle \sigma _{I}^{z}$ & $\displaystyle \sigma _{I}^{z}$ & -$\displaystyle \sigma _{I}^{z}$ & -$\displaystyle \sigma _{L/2 +1-I}^{z}$ & $\displaystyle \sigma _{I+1}^{z}$ \\ \addlinespace
 \hline \addlinespace
  $\displaystyle \tilde\sigma _{I}^{x}$ & $\displaystyle \tilde\sigma _{I}^{x}$ & $\displaystyle \tilde\sigma _{I}^{x}$ & $\displaystyle \tilde\sigma _{I}^{x}$ & $\displaystyle \tilde\sigma _{L/2-I}^{x}$ & $\displaystyle \tilde\sigma _{I+1}^{x}$ \\

 $\displaystyle \tilde\sigma _{I}^{y}$ & $\displaystyle \tilde\sigma _{I}^{y}$ & $\displaystyle \tilde\sigma _{I}^{y}$ & -$\displaystyle \tilde\sigma _{I}^{y}$ & -$\displaystyle \tilde\sigma _{L/2-I}^{y}$ & $\displaystyle \tilde\sigma _{I+1}^{y}$ \\

 $\displaystyle \tilde\sigma _{I}^{z}$ & -$\displaystyle \tilde\sigma _{I}^{z}$ & $\displaystyle \tilde\sigma _{I}^{z}$ & -$\displaystyle \tilde\sigma _{I}^{z}$ & -$\displaystyle \tilde\sigma _{L/2-I}^{z}$ & $\displaystyle \tilde\sigma _{I+1}^{z}$ \\ \addlinespace
 \hline \hline
\end{tabular}
        \caption{(Top half) Symmetry transformations of $S^\alpha_i$ spins (for the Hamiltonian in Eq.~\ref{eqn:eq2.1}) and corresponding fermionic operator $\hat{c}_{i}$ (for Hamiltonian in Eq.~\ref{eq_fermionxxz}) defined on the lattice sites $i$ of a 1-D chain of $L$ sites. For open chains, we consider $L$ to be even with odd number of $J_b$ bonds; (lower half) symmetry transformations of $\sigma (\tilde\sigma)$ spin defined on the odd(even) bonds (Fig.~\ref{fig:fig3}) for a chain of size $L$ as defined by Eq.~\ref{eqn:eq2.7} (\ref{eqn:eq2.7pi}).}
        \label{tab:symm_tab_S}
\end{table*}

\subsection{\texorpdfstring{IPM$_0$}{}}

We start by considering the latter when $J_b\rightarrow\infty$ \textit{i.e.,} all even bonds (between site $2i$ and $2i+1$) are erased such that only the blue bonds remain in Fig.~\ref{fig:fig3} and the Hamiltonian (Eq.~\ref{eqn:eq2.1}) is given by
\begin{align}
    H_b=J_b\sum_i S^z_{2i-1}S^z_{2i}.
    \label{eqn:eq2.3}
\end{align}
The ground state manifold is now $2^L$-fold degenerate ($L$ is the number of sites) and is comprised of two states on each odd-bond $(2i-1,2i)\equiv I$, given by anti-parallel Ising configuration of spins
\begin{align}
    \{|\uparrow\downarrow\rangle, |\downarrow \uparrow\rangle\}_{2i-1,2i}\equiv\{|+\rangle,|-\rangle\}_I.
    \label{eq_GSising}
\end{align}
The excited states comprises of parallel configuration at each bond and has a gap of $J_b/2$. The ground-state manifold of Eq.~\ref{eq_GSising} can be characterized by defining Ising bond spins, $\sigma^\alpha_I$, ($\alpha=x,y,z$) such that
\begin{align}
    \sigma_I^z|\pm\rangle_I=\pm|\pm\rangle_I,
\end{align}
and $\sigma^x_I$ and $\sigma_I^y$ denote, respectively, the symmetric and antisymmetric tunnelling. In terms of operators,
\begin{equation}\label{eqn:eq2.7}
\sigma _{I}^{z} =P^{(2i-1,2i)}_{J_{b}}\left[ S_{2i-1}^{z} -S_{2i}^{z}\right] P^{(2i-1,2i)}_{J_{b}} ,
\end{equation}
where $P^{(2i-1,2i)}_{J_b}$ is projector to ground state of $H_b$ (Eq.~\ref{eqn:eq2.3}) for the bond $I$ 
\begin{equation}\label{eqn:eq2.8}
P^{(2i-1,2i)}_{J_{b}}=|+_I\rangle \langle+_I|+ |-_I\rangle\langle-_I|.
\end{equation}

The action of the symmetries of the system (see above)  on the $\sigma^\alpha_I$ is shown in Table~\ref{tab:symm_tab_S}. It is immediately clear that they correspond to non-Kramers doublets since only $\sigma_I^z$ is odd under TR~\cite{PhysRevB.102.235124}. The terms $J_a$ and $J_\perp$ lift the macroscopic degeneracy, albeit in two different ways.

Treating $J_a$ and $J_\perp$ perturbatively, we can now obtain the low-energy effective Hamiltonian within regular degenerate perturbation theory, which, to the leading order in the two couplings, is given by (see details in Appendix~\ref{appen_degenpert})
\begin{equation}\label{eqn:eq2.15}
\tilde{H}_p [J_b\rightarrow \infty]= -\frac{J_a}{4} \sum_{I}\sigma_I^z\sigma_{I+1}^z -\frac{J_\perp}{2}\sum_I\sigma_I^x.
\end{equation}
Thus, while Ising interactions given by $J_a$ generate ferromagnetic coupling between adjacent bond-spins, the $XY$ terms, proportional to $J_\perp$, act as a transverse field on each bond-spin. These effects are easy to understand from the respective structure of the terms in Eq.~\ref{eqn:eq2.1}. The Ising ($\propto J_a$) term favours 
anti-parallel configuration even on the even (red in Fig.~\ref{fig:fig3}) bonds and this is effectively achieved by aligning the (odd)-bond spins, $\sigma^z_I$. On the other hand, the $XY$ interactions ($\propto J_\perp$) on every even bond want to switch $|\uparrow\downarrow\rangle\leftrightarrow|\downarrow\uparrow\rangle$ and is
therefore a transverse field for the bond spins. Thus the leading order effective Hamiltonian (Eq.~\ref{eqn:eq2.15}) is describes a nearest-neighbour transverse-field-ferromagnetic Ising chain, which can be exactly solved to obtain two phases as a function of $J_\perp/J_a$ with the phase transition occurring at 
\begin{equation}
J_a ~=~ 2J_\perp \label{pt1} \end{equation} 
according to the KW self-duality of the transverse field Ising chain. 

In Appendix~\ref{appen_degenpert}, we have also extended the result
in Eq.~\ref{pt1} to first order in an expansion in $1/J_b$. For
large $J_b$, we find a more accurate expression,
\begin{equation}
J_a ~=~ 2J_\perp ~-~ \frac{J_\perp^2}{2 J_b}, \label{pt2} \end{equation} 
for the phase transition line which separates IPM$_0$ 
and IN in Fig.~\ref{fig:pd_cc}.

For $J_a\gg J_\perp$, we have ferromagnetic ordering in terms of the bond spins which spontaneously break the global $Z_2$ symmetry for the bond spins. This is exactly the time-reversal symmetry. In addition, it also breaks the reflection symmetry about the odd bonds, as denoted by $\mathcal{R}_{J_b}$ in Table~\ref{tab:symm_tab_S}. Remarkably, in terms of the underlying, $S_i^z$, spins this is a IN  or the CDW phase (Fig.~\ref{fig:pd}) in terms of the fermions or bosons in Hamiltonian of Eq.~\ref{eq_fermionxxz} with the latter representation used for numerical calculations here as well as in Ref.~\cite{PhysRevB.106.L201106}. On the other hand, for $J_a\ll J_\perp$, the $Z_2$ symmetric paramagnet is obtained. This, is the trivial phase IPM$_0$ where the two spins on the odd bond $I$ form a $XY$-singlet (Fig.~\ref{fig:pd}). Both these phases are gapped and are separated by an intervening continuous phase transition belonging to the Ising universality class that is described by a chiral $c=\frac{1}{2}$ (1+1)CFT. 

The phase transition is best captured in terms of fermions, $d_I$, obtained via a JW transformation that is similar to Eq.~\ref{eqn:eq5}, albeit for the bond spins, $\sigma^\alpha_I$. The phase diagram in terms of the fermions is sketched schematically in Fig.~\ref{fig:QIM_PD}. 

\begin{figure}
    \centering
\includegraphics[width=1\columnwidth]{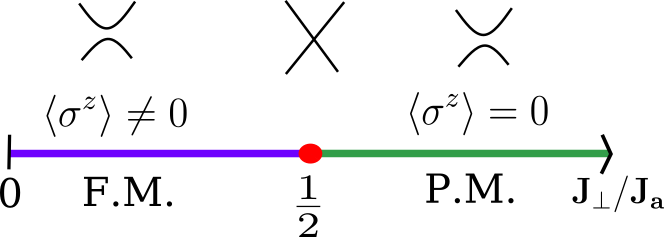}
    \caption{Phase diagram of a quantum Ising model at T=0. At $J_\perp / J_a =1/2$ there is a quantum phase transition from a ferromagnetically  ordered phase to a quantum paramagnet.}
    \label{fig:QIM_PD}
\end{figure}

\subsection{\texorpdfstring{IPM$_\pi$}{}}

Now consider starting with Eq.~\ref{eqn:eq2.1} and taking the $J_a\rightarrow\infty$ \textit{i.e.}, all even (red) bonds (between site $2i-1$ and $2i$) are present (Fig.~\ref{fig:fig3} and also Fig.~\ref{fig:top_chain}) such that Hamiltonian is given by $H_a= J_a\sum_i S_{2i}^z S_{2i+1}^z.$ Similar to IPM$_0$ we now get bond spins on the even bonds, $\tilde\sigma^\alpha_{\tilde I}$ following the same construction, albeit on the even bonds with
\begin{align}
    \tilde\sigma_{\tilde I}^z = P^{(2i,2i+1)}_{J_{a}}\left[ S_{2i}^{z} -S_{2i+1}^{z}\right] P^{(2i,2i+1)}_{J_{a}}.
    \label{eqn:eq2.7pi}
\end{align}

There is, however, a crucial difference that in an open system the two end spins are now completely decoupled, as shown in Fig.~\ref{fig:top_chain}, which leads to two dangling edge zero modes. Leaving aside the edge modes for the time being, a perturbation theory in $J_b$ and $J_\perp$ for the bulk gives rise to a transverse ferromagnetic field Ising model for the $\tilde\sigma_{\tilde I}$ bond spins similar to Eq.~\ref{eqn:eq2.15}. In particular the leading order Hamiltonian is now given by 
\begin{equation}\label{eqn:eq2.19b}
\tilde{H}_p [J_a\rightarrow \infty]= -\frac{J_b}{4} \sum_{\tilde I}\tilde\sigma_{\tilde I}^z \tilde\sigma_{\tilde I+1}^z -\frac{J_\perp}{2}\sum_{\tilde I} \tilde\sigma_{\tilde I}^x,
\end{equation}
which supports a $Z_2$ symmetry broken IN phase for $J_b\gg J_\perp$ and a $Z_2$ symmetric IPM in the opposite limit with an intervening Ising phase transition.

However, due to the dangling edge spins the paramagnetic phase is non-trivial and in fact is an SPT similar to the Haldane phase. To understand this, consider the limit of $J_b=J_\perp=0$ whence the presence of the dangling zero energy edge spins are clear (Fig.~\ref{fig:top_chain}). It is noteworthy that each spin-$\frac{1}{2}$ at a given site satisfies $\mathcal{T}_i^2=-1$. However, in the bulk, each bond-spin is actually a non-Kramers doublet such that $\mcl{T}^2_{\tilde I}=+1~~(\forall~\tilde I)$. This makes the edge modes to be robust to weak time-reversal invariant perturbations such as $J_\perp, \ J_b$ as long as the bulk gap remains intact. This leads to 4-fold degenerate edge modes where the degeneracy can be lifted by perturbations such as an external magnetic field, which breaks the time-reversal symmetry. 

\begin{figure}
    \centering
    
\includegraphics[width=1\columnwidth]{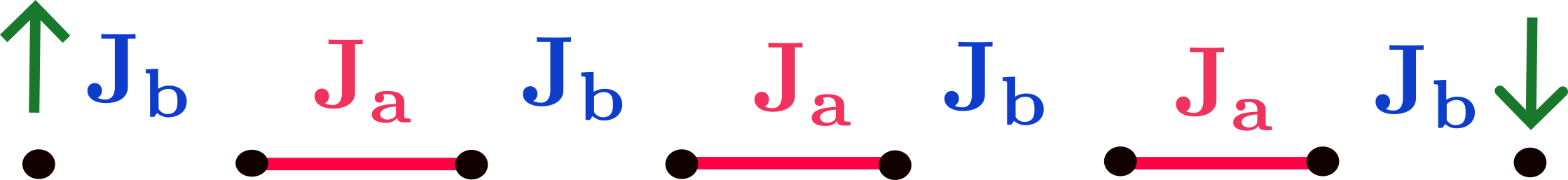}
    \caption{ For an open chain, in the $J_a\rightarrow\infty$ limit, we have two free edge spin-$\frac{1}{2}$'s
at the boundary
which leads to a four-fold degeneracy of the ground state.}
    \label{fig:top_chain}
\end{figure}

Thus, the IPM$_\pi$ is a symmetry-protected topological (SPT) phase distinct from IPM$_0$ and this necessitates the gapless TLL across which the topological invariant changes. It is also due to this, the single (fermion) particle gap must close across the IN-IPM$_\pi$ transition, unlike the IN-IPM$_0$ transition even though both belong to the same (Ising) universality class. Further details about the SPT is discussed in Ref.~\cite{SPT_Wen, chen2012symmetry}.

\subsection{SPT via decorated domain walls}
\label{sec_ddw}

\begin{figure}
    \centering
    \includegraphics[scale=0.5]{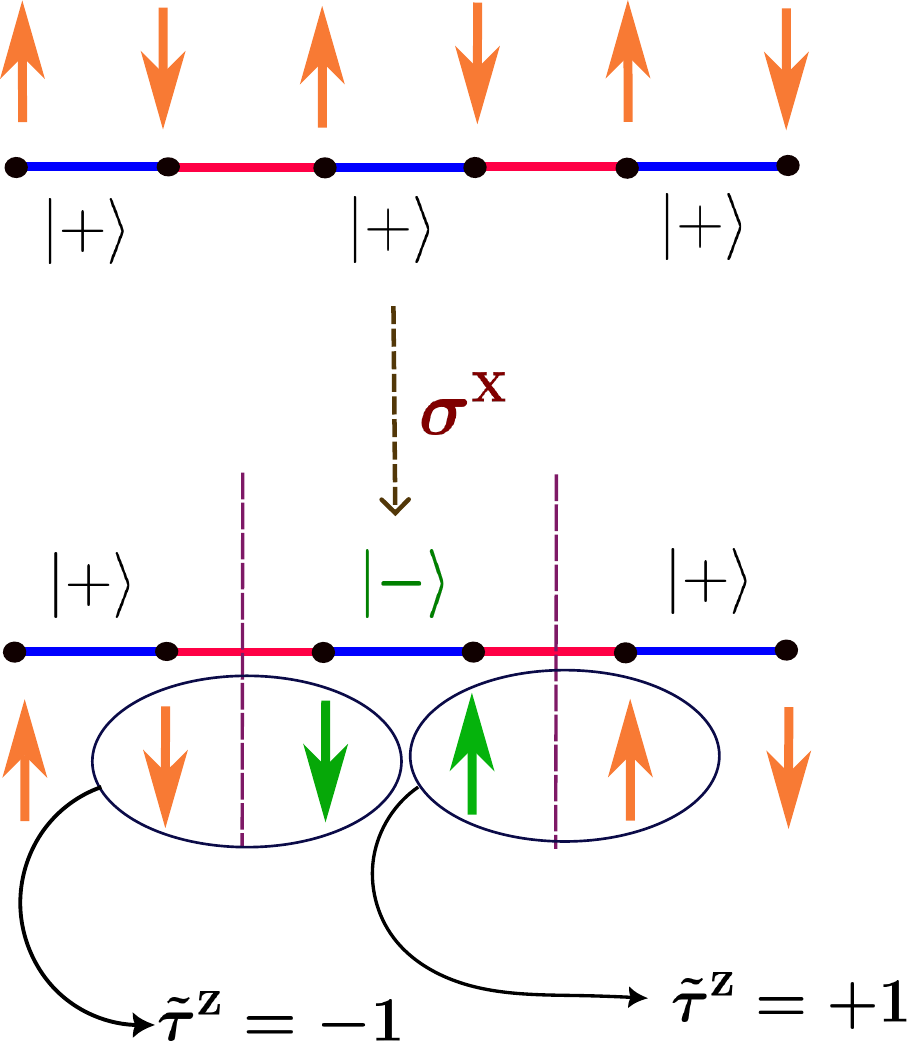}
    \caption{Domain wall in the IN phase for $J_b>J_a$. Note that the transverse field term in Eq.~\ref{eqn:eq2.15}, which creates a domain wall in the ordered phase for the $\sigma_I^z$ (odd) bond spins, automatically leads to the excited state doublet $\tilde\tau^z_{\tilde I}=\pm 1$ on the domain walls which are even bonds of the original lattice (see text).}
    \label{fig:domainwallspt}
\end{figure}

Useful microscopic insights of the above SPT are obtained using the decorated domain wall picture following Ref.~\cite{chen2014symmetry}. This is best done in the anisotropic limits discussed above and is summarized in Fig.~\ref{fig:domainwallspt}. To be precise, let us start from the limit $J_b\rightarrow\infty$. As discussed above, the low-energy degrees of freedom are the Ising spins, $\sigma^\alpha_I$ (Eq.~\ref{eqn:eq2.7}), defined on the odd bonds. Then turning on $J_a$ leads to the Ising order for the bond spins via the first term in Eq.~\ref{eqn:eq2.15} with one of the ground states shown in the top panel of Fig.~\ref{fig:domainwallspt} where $\sigma^z_I=+1~~\forall~I$. Notice that for such a configuration the even bonds are in the low-energy state $\tilde\sigma^z_{\tilde I}=-1$ (Eq.~\ref{eqn:eq2.7pi}). Now, creating a domain wall (by the transverse field-- the second term in Eq.~\ref{eqn:eq2.15}) in this ordered configuration, as shown in Fig.~\ref{fig:domainwallspt}, cost an energy $2J_a$ and the domain wall sits on the odd bonds adjacent to the flipped spin. However, if we go back to the underlying $S^\alpha_i$ spins, we notice that (lower panel of Fig.~\ref{fig:domainwallspt}) on the odd bonds on which the domain wall sits, both $S^\alpha_i$ spins are parallel and therefore do not lie in the manifold described by $\tilde\sigma^\alpha$. Instead they are in the high energy manifold of $J_aS^z_{2i}S^z_{2i+1}$ which consists of $|\uparrow\uparrow\rangle, |\downarrow\downarrow\rangle$. Ascribing another Ising doublet, $\tilde\tau_{\tilde I}^\alpha$ to describe these doublets on the even bonds (as indicated in Fig.~\ref{fig:domainwallspt}), we see that the domain walls in $\sigma_I^z$ are always attached with $\tilde\tau^z_{\tilde I}=\pm$ spins.  On proliferating such domain walls, the IN order is lost and we have an IPM. However, for $J_a>J_b$, {\it i.e.}, in IPM$_\pi$, the domain wall can hit the boundary and this leads to the edge mode while such domain walls are restricted to the bulk for IPM$_0$. The above picture can be made concrete via duality as discussed in Appendix~\ref{app:XXZ_to_QAT}.


\section{The mean-field phase diagram}
\label{sec_mft}

Having discussed the SPTs, we now turn to the MF phase diagram. To this end, we use the well-known fermionic representation of the spin-Hamiltonian (Eq.~\ref{eqn:eq2.1}) obtained via the Jordan-Wigner transformation. The details of the spin-fermion mapping are well known and are summarized in Appendix~\ref{appen_spin-fermion} for completeness. The resultant fermionic Hamiltonian  (up to an overall constant) is given by
\begin{equation}
 \begin{split}
       H=& -\frac{J_\perp}{2}\sum_{i}\left(c_i^\dag c_{i+1} +  \text{H.c.} \right)-\frac{J_a+J_b}{2}\sum_{i} \hat{n}_i \\
       &+ J_b\sum_{i}\hat{n}_{2i-1}\hat{n}_{2i} + J_a\sum_{i} \hat{n}_{2i} \hat{n}_{2i+1},
         \label{eq_fermionxxz}
   \end{split}
\end{equation}
which at half filling forms a faithful representation of the XXZ model. 

To obtain the MF description we consider the decomposition of the density-density interactions in three different channels: the bond fields (on the odd and the even bonds that capture IPM$_0$ and IPM$_\pi$) and the Ising Neel (CDW for fermions) order parameter. The former is characterized by 
\begin{equation}\label{eqn:eq2.22}
m_o = \frac{1}{L}\sum_{i}\langle c_{2i-1}^\dag c_{2i}+ c_{2i}^\dag c_{2i-1}\rangle
\end{equation}
for the odd bonds, and 
\begin{equation}\label{eqn:eq2.23}
m_e = \frac{1}{L}\sum_{i}\langle c_{2i}^\dag c_{2i+1}+ c_{2i+1}^\dag c_{2i}\rangle
\end{equation}
for the even bonds. It is easy to see, using Eqs.~\ref{eqn:eq5},~\ref{eqn:eq2.7}  and~\ref{eqn:eq2.7pi} that in the anisotropic limit of $J_b (J_a)\rightarrow\infty$, the bonding orbital on odd (even) bond, {\it i.e.}, $m_o\sim \sum_I\langle\sigma_I^x\rangle ~(m_e\sim \sum_{\tilde I}\langle\tilde\sigma^x_{\tilde I}\rangle)$.

On the other hand, the IN or CDW order parameter is given by
\begin{align}
    n_-=\frac{1}{L}\sum_i \frac{\langle \hat{n}_{2i}-\hat{n}_{2i+1}\rangle}{2}=\frac{1}{2}\left(n_A-n_B\right),
\end{align}
where $n_A$ and $n_B$ are fermion densities on even and odd sites respectively. Therefore, in the anisotropic limit we have $n_-$  proportional to $\sim \sum_I\langle\sigma^z_I\rangle~(\sum_{\tilde I}\langle\tilde\sigma^z_{\tilde I}\rangle)$ in the limit of $J_b (J_a)\rightarrow\infty$. The above identification is explicit from the symmetry transformation of $m_o, m_e$ and $n_-$ Table~\ref{tab:symm_tab_ops} (Appendix~\ref{appen_mftdetails}) when compared with Table~\ref{tab:symm_tab_S}.

\begin{figure}
        \centering
        \includegraphics[width=1\columnwidth]{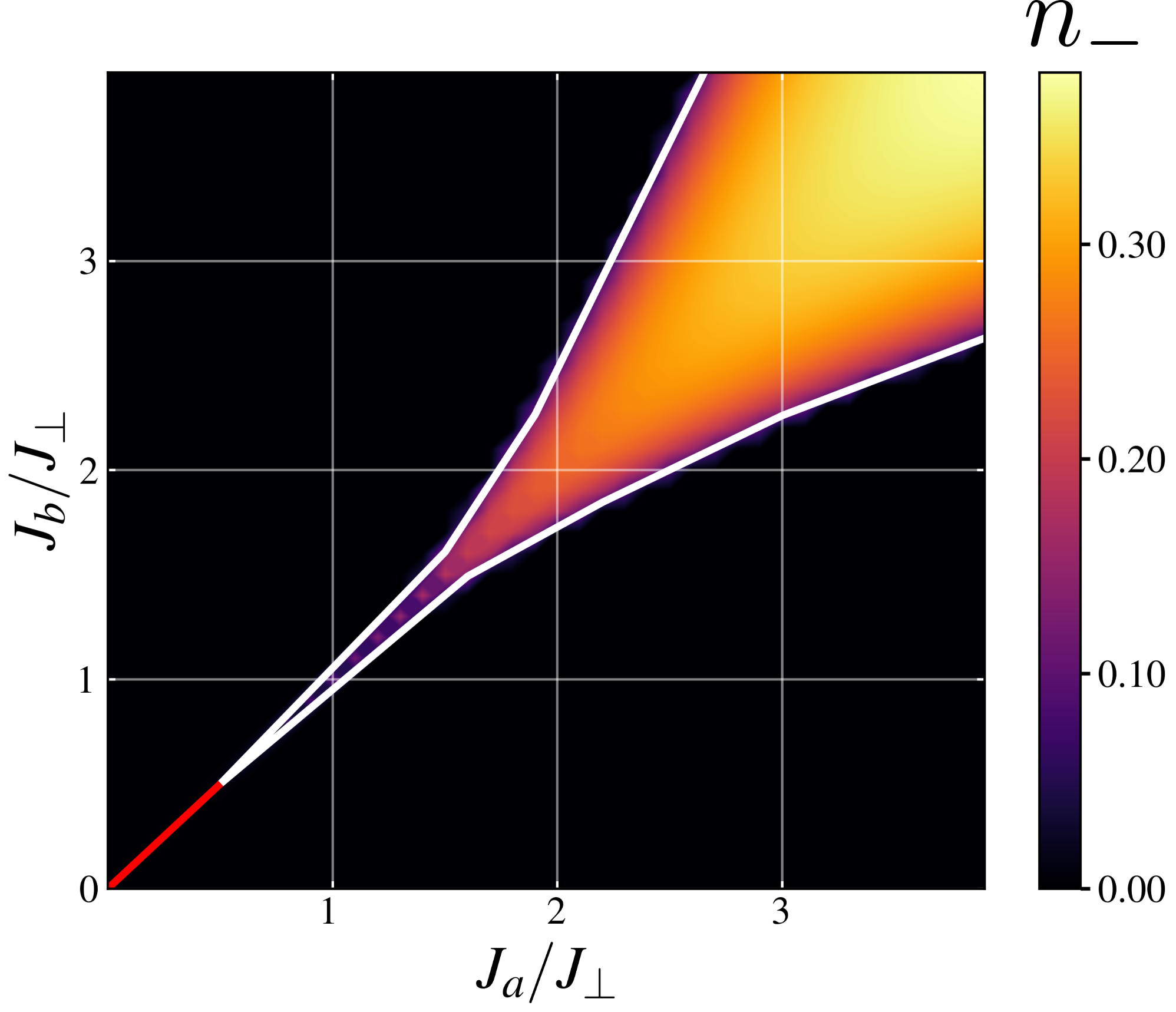}
\caption{The MF phase diagram obtained via tracking the order parameter $n_-$ and single-particle gap for different values of $(J_a, J_b)$ (Fig.~\ref{fig:mftpara} upper and lower panels) that obtained through minimization of MF energy (Eq.~\ref{eq:GS_Energy_full_MFT_comb}) for $\alpha=0.6$. The white line indicates the contour when $n_-$ becomes non-zero while the red line distinguishes the two IPM phases across the $J_a=J_b$ line whence the single-particle gap is zero. The qualitative similarity with the numerically obtained phase diagram (Fig.~\ref{fig:pd}
    ) is evident.}
    \label{fig:MFT_PD_comb_nm}
\end{figure}

 To decouple the density-density interactions in the above three channels, we note that each bond contains one of the two bond channels and the CDW channel. Further, we use a phenomenological weight parameter $\alpha \in [0, 1]$ to bias one of the two channels to understand the generic nature of the phase diagram. This leads to the MF decoupling in terms of $n_+=(n_A+n_B)/2, \ n_-$, $m_o$, and $m_e$ :

\begin{equation} \label{MF_Jb}
    \begin{split}
        &J_b\hat{n}_{2i-1}\hat{n}_{2i}\rightarrow \\
        & J_b (1-\alpha)\Big[ \frac{\hat{n}_i}{2} -m_o (c_{2i-1}^\dag c_{2i}+ c_{2i}^\dag c_{2i-1})  +\frac{m_o^2}{2}  \Big] \\
        & + J_b \alpha \Big[  n_-^2 - n_+^2   +(n_+ - n_-)\hat{n}_{2i} +(n_+ + n_-)\hat{n}_{2i+1} \Big]
    \end{split}
\end{equation}
for the odd bonds, and
\begin{equation} \label{MF_Ja}
    \begin{split}
        &J_a\hat{n}_{2i}\hat{n}_{2i+1}\rightarrow \\
        & J_a (1-\alpha)\Big[ \frac{\hat{n}_i}{2} -m_e (c_{2i}^\dag c_{2i+1}+ c_{2i+1}^\dag c_{2i})  +\frac{m_e^2}{2}  \Big] \\
        & + J_a \alpha \Big[  n_-^2 - n_+^2   +(n_+ - n_-)\hat{n}_{2i} +(n_+ + n_-)\hat{n}_{2i+1} \Big].
    \end{split}
\end{equation}

\begin{figure}
    \centering
    \includegraphics[width=0.9\columnwidth]{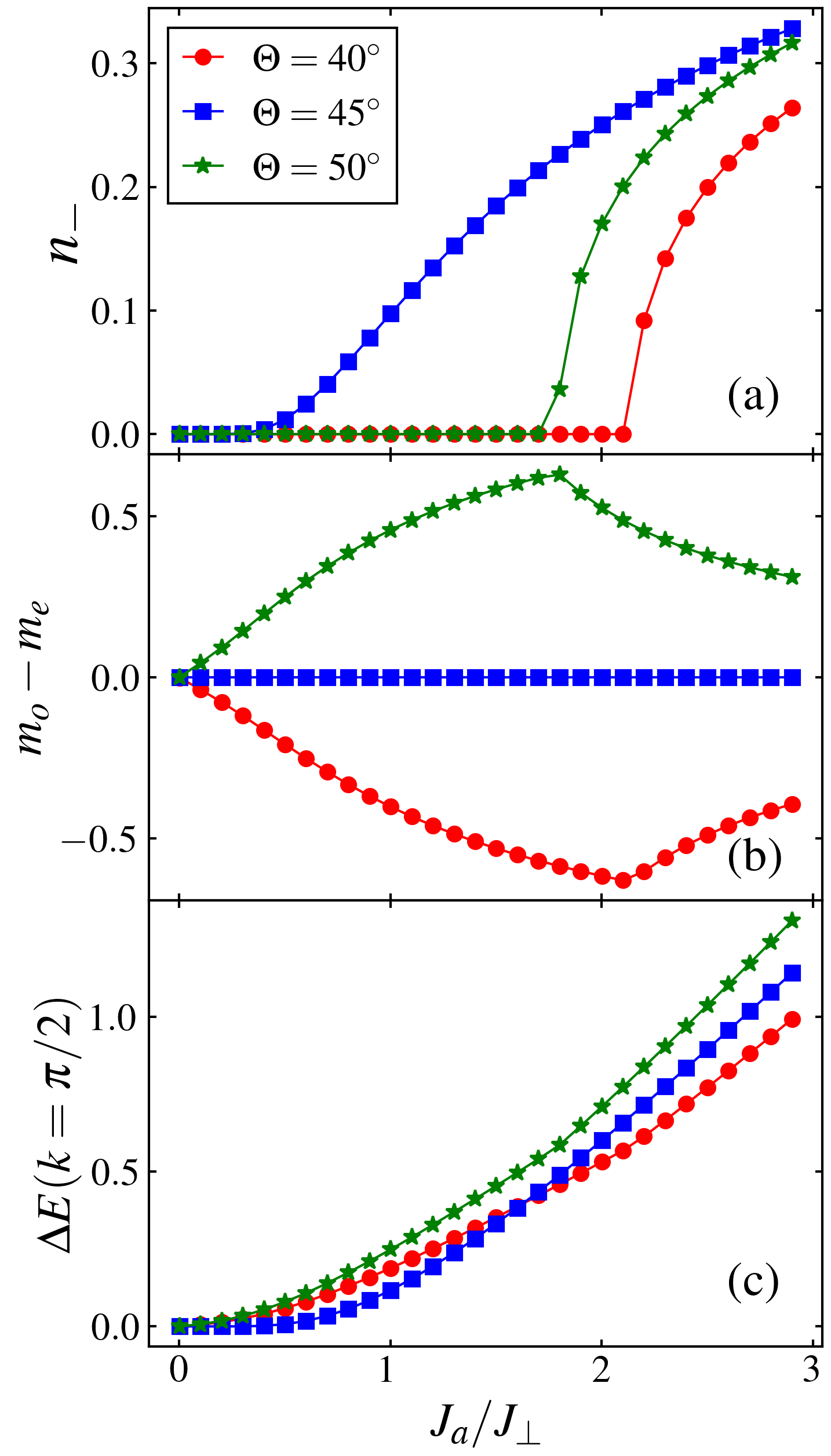}
    \caption{The MF parameters for  selected cuts at different $\Theta=\tan^{-1}(J_b/J_a)$ and $\alpha=0.6$. We have plotted (a) the CDW order parameter $n_-$, (b) the difference of the dimerization fields, {\it i.e.}, $m_o-m_e$ which is zero for the symmetric case $\Theta=\pi/4$ and positive (negative) for $\Theta<(>)\pi/4$, (c) minimum energy gap $\Delta E(k=\pi/2)$.}
    \label{fig:mftpara}
\end{figure}

The resultant MF Hamiltonian can be solved self-consistently with respect to  $n_+$, $n_-$, $m_o$, $m_e$ and gives rise to the phase diagram given in Fig.~\ref{fig:MFT_PD_comb_nm} with the details of the values of various MF parameters for selected cuts in Fig.~\ref{fig:mftpara} (further details are given in Appendix~\ref{appen_mftdetails}). Note that for $\alpha\leq0.5$, $n_-$ exhibits a zero value across all regions except those situated beyond a certain point along a 45-degree line.  However, when $\alpha$ exceeds 0.5, a discernible departure occurs, and $n_-$ takes on non-zero values extending beyond a critical point as we traverse along any angular direction.

The self-consistently determined IN order parameter that signals the TR symmetry-breaking is plotted in the upper panel of Fig.~\ref{fig:mftpara} for $\alpha=0.6$. Note that for $\alpha \leq 0.5$, $n_-$ remains zero across all regions except beyond a specific point along a 45-degree line (Fig.~\ref{fig:MFT_PD_comb_nm_Appendix}(a)). However, as $\alpha$ exceeds 0.5, a noticeable shift occurs, and $n_-$ adopts non-zero values, extending beyond a critical point as we traverse along any angular direction (Figs.~\ref{fig:MFT_PD_comb_nm} and~\ref{fig:MFT_PD_comb_nm_Appendix}(b)).

In Fig.~\ref{fig:mftpara}, the IN order parameter $ n_-$ is shown for different cuts in $J_a-J_b$ plane (top panel), the difference of the bond fields, $m_o-m_e$, in the middle panel and the minimum of the single-particle excitation gap in lower panel. Not surprisingly, $m_o=m_e$ along the $J_a=J_b$ line whence the single-particle gap closes at $\pi/2$ for $n_-=0$. 

\begin{figure}
    \centering   
        \includegraphics[width=0.9\linewidth]{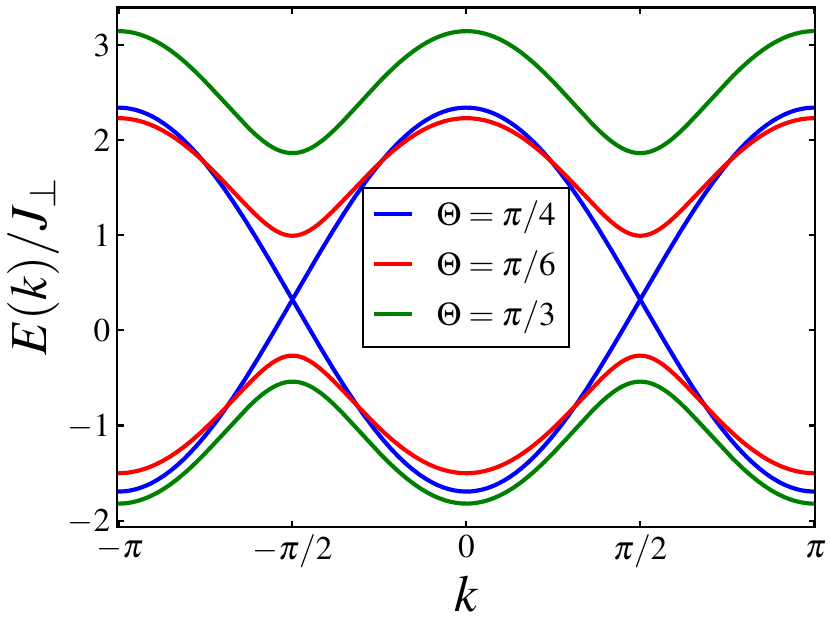}%
        \label{fig:14a}%

    \caption{Band structure for values of $ J_b, \ m_o, \ m_e$ along $\pi/6$, $\pi/4$ and $\pi/3$ lines in $J_a-J_b$ plane for $J_a = 0.8$. For $\theta = \pi/4$ or $J_a=J_b$ case spectrum is gapless, while for the other two cases, it is gaped.}\label{fig14}
\end{figure}

In the region where $n_-=0$, the MF reduces to the two bond parameters $m_e$ and $m_o$. The self-consistent fermion band-structure for $\alpha=0$ (see Appendix~\ref{appen_mftdetails}) is shown in Fig.~\ref{fig14} for various points in the $J_a-J_b$ plane. In this region, the mean-field Hamiltonian is given by (for $\alpha=0$)
\begin{equation}
    \begin{split}
        \label{eqn:eq4.15}
           H_{M.F.}(\alpha=0) = &
           -(\frac{J_\perp}{2} + J_b m_o )\sum_{i=1}^{L} \left[ c_{2i}^{\dag} c_{2i-1}+ c_{2i-1}^\dag c_{2i}\right]\\
           &- (\frac{J_\perp}{2} + J_a m_e )\sum_{i=1}^{L-1} \left[ c_{2i}^\dag c_{2i+1}+ c_{2i+1}^\dag c_{2i}\right],
    \end{split}
\end{equation}
where $2L$ is the total number of sites for an open chain. Eq.~\ref{eqn:eq4.15} is 
precisely the SSH model which shows a phase transition between the topological to trivial insulating phases corresponding to IPM$_\pi$ and IMP$_0$ respectively with single-particle gap closing at momentum $\pm\pi/2$ (Fig.~\ref{fig14}) for $J_a=J_b$ corresponding to the red line in Fig.~\ref{fig:MFT_PD_comb_nm}. Solving it numerically for $2L=100$ with open boundary condition, we get zero energy modes for values of $J_a, \ J_b, \ m_o, \ m_e$ along $\theta=\pi/6$ line ($J_b= J_a \tan\theta$) or in general for $J_b/J_a <1$ while no zero energy modes for $J_b/J_a >1$. Thus, these two phases are topologically distinct at the mean-field level.

\section{The Phase transitions: Numerical results}
\label{sec_dmrg}

Having understood the phases, we now move on to the phase transitions. The numerically calculated central charge (Fig.~\ref{fig:pd_cc}; details in Appendix~\ref{appen_centralcharge}) identifies the nature of the three transitions. This is best understood starting with the $c=1$ a multicritical point for $J_a=J_b=1$ which is just the critical (Heisenberg) point of the undimerized XXZ model. The two $c=\frac{1}{2}$ Ising critical lines emerge from this multicritical point and describe the transition between the two IPMs and the IN. We now turn to investigate these transitions, starting with our numerical results and MF theory, followed by the critical theory using bosonization approaches. All the numerical calculations are performed using DMRG method under open boundary condition (OBC), unless otherwise explicitly mentioned.

 \begin{figure}[!t]
\centering
\includegraphics[width=1\columnwidth]{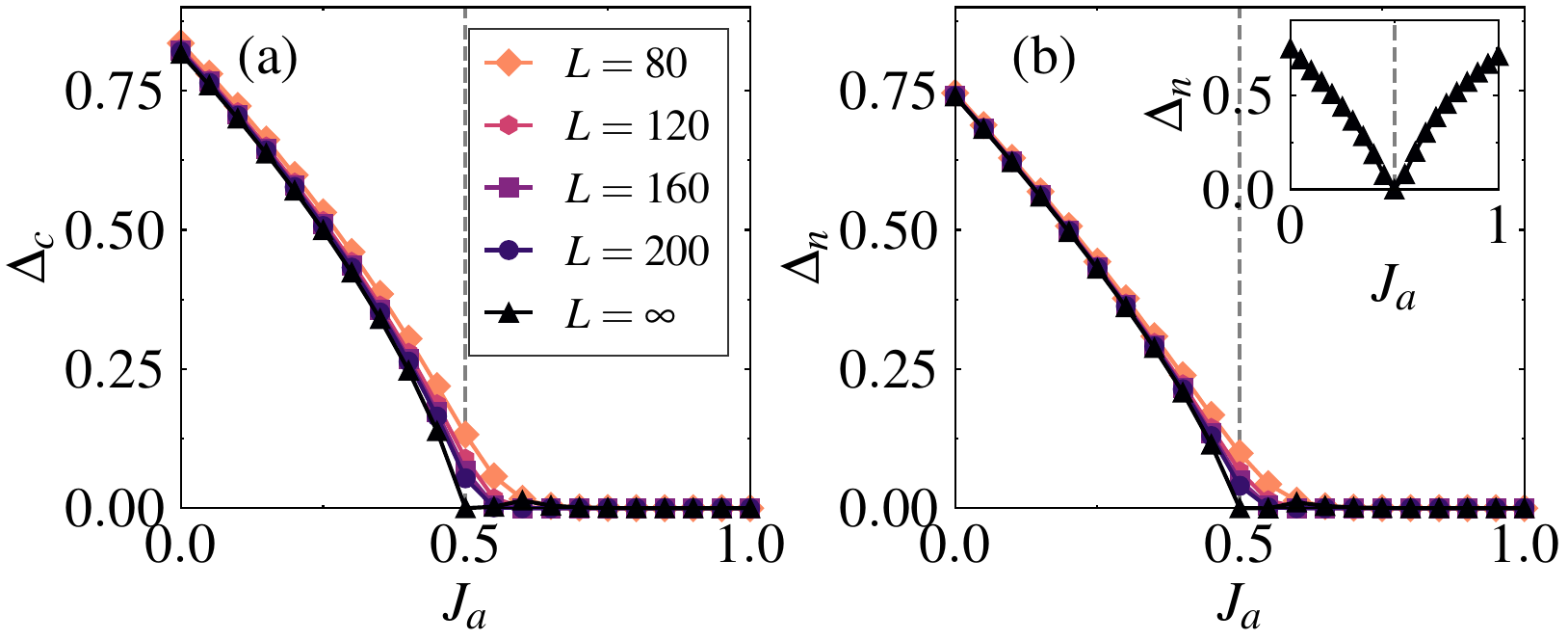}
\caption{(a) Charge gap $\Delta_c$ and (b) neutral gap $\Delta_n$ as a function of $J_a$ and varying system sizes for $J_b=0.5$. The dashed grey lines mark the critical point corresponding to the IPM\texorpdfstring{$_0$}{}-IPM\texorpdfstring{$_{\pi}$}{} phase transition, \textit{i.e.,} $J_a=0.5$. The inset of (b) shows $\Delta_n$ for the system under PBC at $L=\infty$ by extrapolating the data with $L=40, 60, 80$ and $100$.}
\label{fig:tpt_gaps}
\end{figure}
\subsection{\label{sec:dmrgtpt} IPM\texorpdfstring{$_0$}{} to IPM\texorpdfstring{$_{\pi}$}{} phase transition}

We start by discussing the transition between the two IPMs across the TLL. For this purpose we consider a cut through the phase diagram for $J_b=0.5$ and varying $J_a$. Note that in this case, $J_a=0.5$ is the critical point. The criticality of the phase transition can be captured by the closing of the single fermion gap defined as
\begin{align}
\Delta_c=E_0(N+1)+E_0(N-1)-2E_0(N),
\label{eq:ng}
\end{align}
where $E_0(N)$ is the ground state energy for $N$ total number of particles in the system. In Fig.~\ref{fig:tpt_gaps}(a) we plot $\Delta_c$ as a function of $J_a$ for varying system sizes as well as at the finite size extrapolated value in the thermodynamic limit ($L=\infty$). The extrapolated value is extracted using a quadratic fitting of $\Delta_c$ versus $1/L$. Clearly, the gap $\Delta_c$ is finite in the trivial IPM$_0$ phase ($J_b<1$) and vanishes in the topological IPM$_\pi$ phase ($J_b>1$) and, of course, at the critical point. This means that creation of a particle or a hole at half-filling requires an extra amount of energy in the trivial phase. However, in the topological IPM phase, $\Delta_c$ vanishes due to the existence of the zero energy edge states for which the energy cost for adding and removing a particle at the edges is zero in the thermodynamic limit.  
 
Insights can also be obtained from the neutral gap defined by the formula
\begin{align}
\Delta_n=E_1(N)-E_0(N),
\label{eq:ng2}
\end{align}
where $E_0(N)$ is the ground state energy and $E_1(N)$ is the first excited state energy for $N$ number of particles~\cite{Ejima_2014, Melo_2023}. In Fig.~\ref{fig:tpt_gaps} (b) we plot $\Delta_n$ for the same set of parameters considered in Fig.~\ref{fig:tpt_gaps}(a). We obtain that $\Delta_n$ remains finite in the trivial IPM phase but vanishes at the critical point and inside the topological IPM phase. The reason behind this can be attributed to the particle distributions in the two IPM phases, which can be understood as follows. The ground state of the trivial IPM phase is unique, \textit{i.e.,} one particle is dimerized in every odd bond of the lattice. However, the ground state of the topological IPM phase exactly at half-filling is two-fold degenerate for open boundary conditions due to the presence of the edge states. In the bulk, one particle is dimerized in every even bond of the lattice but one of the edges (either left or right sites) is filled and the other is empty. Thus, two configurations are possible depending on which edge is filled, and the ground state becomes degenerate. The degeneracy gets lifted when PBC is assumed, where the concept of edge states is irrelevant and one cannot distinguish both the IPM phases. As expected, the extrapolated values of $\Delta_n$ in the thermodynamic limit ($L=\infty$) vanishes only at the critical point, and becomes finite on either sides of it (see the inset of Fig.~\ref{fig:tpt_gaps}(b)).

\begin{figure}
\centering
\includegraphics[width=0.9\columnwidth]{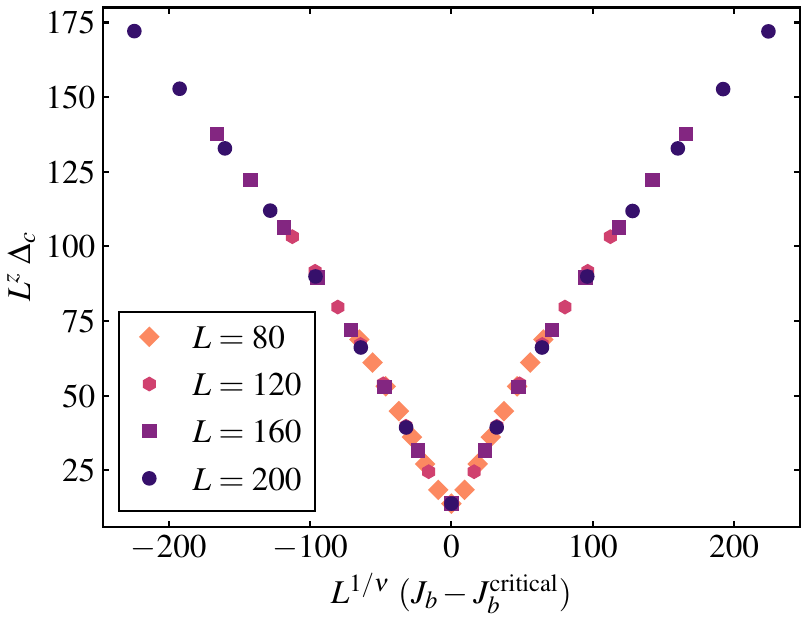}
\caption{Data collapse of the charge gap $\Delta_c$ computed for different system sizes $L$ and for a cut at $J_a = 0.75$ (along $135^0$ line in $J_a - J_b$ plane). Here, correlation length exponent $\nu\sim0.74$ for this cut and dynamical critical exponent $z=1$. The critical point $J_b^{\text{critical}}$
  in this case is 0.75, as the critical line is defined by $J_a = J_b$ (Fig.~\ref{fig:pd_cc}).}
\label{fig:cur_collapse_gaps}
\end{figure}

To compare the results of the DMRG and Bosonization calculations (Sec.~\ref{sec_bosonize}) we do a finite size scaling analysis of charge gap $\Delta_c$ by plotting $L^z\Delta_c$ versus $L^{1/\nu}(J_b-J_b^{\text{critical}})$ as shown in Fig.~\ref{fig:cur_collapse_gaps}. The charge gap was computed for various system sizes (under PBC), revealing a curve collapse consistent with the correlation length exponent $\nu=1/(2-K)$ (which is $\sim0.74$ for $J_a = 0.75$ cut), as predicted by Bosonization (Eq.~\ref{eqn:total_betat}), and a dynamical critical exponent $z=1$. To obtain the value of $K$ used in $\nu$ we employed Eq.~\ref{eq:total_Bose_Ham_paramt} for $J_a=J_b=0.75,~J_\perp=1$. The strong curve collapse indicates that the DMRG data is consistent with the critical exponents predicted by Bosonization.

\subsection{\label{sec:dmrgbocdw} IPM\texorpdfstring{$_0$}{} to IN phase transition}

\begin{figure}
\centering
\includegraphics[width=1\columnwidth]{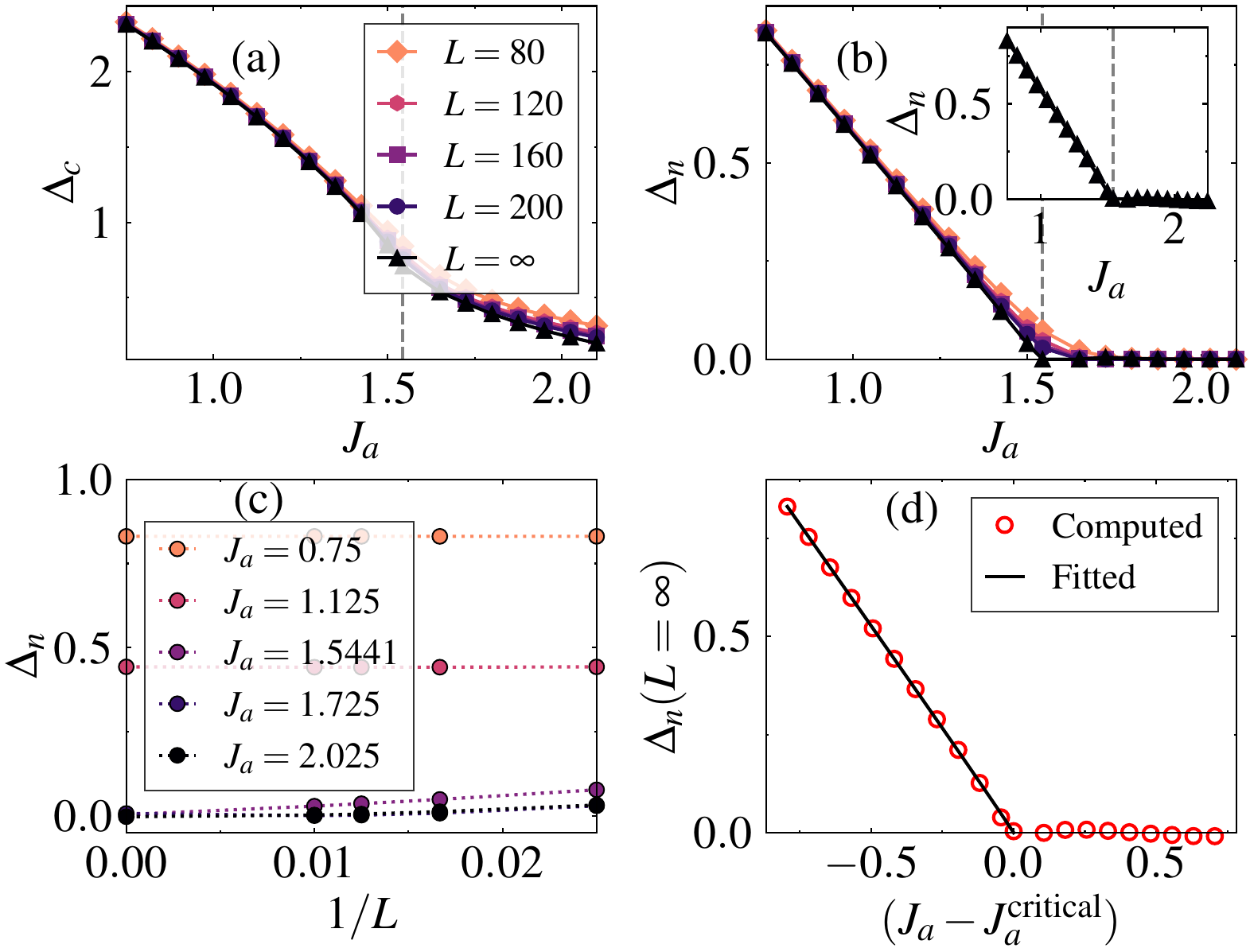}
\caption{(a) Charge gap $\Delta_c$ and (b) neutral gap $\Delta_n$ as a function of $J_a$ for $J_b=1.75$ and varying system sizes. The dashed grey lines mark the critical point corresponding to the IPM$_0$-IN phase transition,\textit{ i.e.,} $J_a=1.5441$. The inset of (b) shows $\Delta_n$ for the system under PBC at $L=\infty$ by extrapolating the data with $L=40, 60, 80$ and $100$. (c) System size dependence of $\Delta_n$ for different values of $J_a$. (d) The red dots represent $\Delta_n$ in the thermodynamic limit $(L=\infty)$ as a function of $(J_a-J_a^{\rm critical})$. The solid black line fits the functional form (see text for more details).}
\label{fig:bocdw_gaps}
\end{figure}

Turning to the transition between the trivial paramagnet (IPM$_0$) and the IN phase, we consider a cut at $J_b=1.75$ of the phase diagram of Fig.~\ref{fig:pd} that marks the transition between the IPM$_0$ and the IN phases. The critical point of the transition between these phases for $J_b=1.75$ was found to be $J_a\sim1.5441$ through a scaling analysis of the structure factor in Ref.~\cite{PhysRevB.106.L201106}. Here we plot the single particle gap $\Delta_c$ as a function of $J_a$ in Fig.~\ref{fig:bocdw_gaps}(a) for different system sizes ($L=80,~120,~160,~200$). We also plot the extrapolated values of the gap $\Delta_c$ in the thermodynamic limit (\textit{i.e.,} for $L=\infty$), which are extracted using a quadratic fitting of $\Delta_c$ as a function of $1/L$. We obtain that $\Delta_c$ remains finite in both the IPM$_0$ and the IN phases and also at the transition point, indicating a continuous phase transition between the two gapped phases even for OBC. On the other hand, the neutral gap 
$\Delta_n$ exhibits a completely different behaviour as a function of $J_a$ which is depicted in Fig.~\ref{fig:bocdw_gaps}(b) for different system sizes as well as the extrapolated values in the limit of $L=\infty$. As mentioned previously, $\Delta_n$ remains finite in the trivial IPM phase due to the unique ground state. At the critical point, \textit{i.e.,} at $J_a \sim 1.5441$, $\Delta_n$ vanishes indicating a phase transition. The apparent gaplessness in the IN phase, \textit{i.e.} for $J_a > 1.5441$, however, is due to the two $Z_2$ symmetry broken ground states that are degenerate in the thermodynamic limit. In the inset of Fig.~\ref{fig:bocdw_gaps}(b), we plot $\Delta_n$ for PBC in the thermodynamic limit  which shows no difference as compared to the OBC case. These analyses based on the charge and neutral gaps clearly identify various quantum phases and transitions between them, which are depicted in Fig.~\ref{fig:pd}. From Fig.~\ref{fig:bocdw_gaps}(c) we further observe that the gap does not close at all in the IPM$_0$ phase, but at the critical point it decreases with increasing system size and closes in the thermodynamic limit. On the other hand, the gap is vanishingly small even for finite system sizes in the IN phase.

 \begin{figure}
\centering
\includegraphics[width=1\columnwidth]{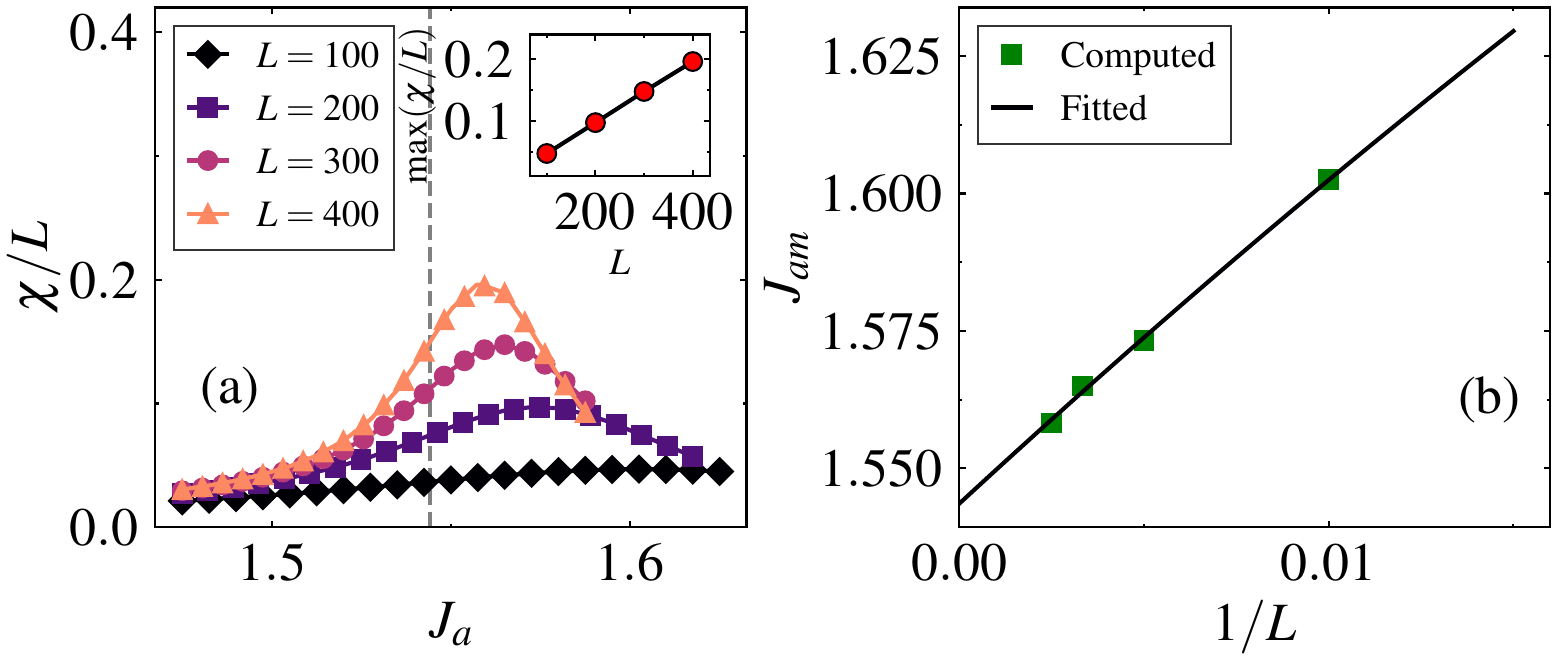}
\caption{(a) Fidelity susceptibility $\chi/L$ as a function of $J_a$ for $J_b=1.75$ and varying system sizes. The dashed grey line marks the critical point corresponding to IPM$_0$-IN phase transition, \textit{i.e.,} $J_a=1.5441$. The inset shows the peak height of $\chi/L$ as a function of system size. (b) The position of the maximum peak height $J_{am}$ is marked by green squares as a function of inverse system size. The black line denotes the corresponding fitted data.}
\label{fig:fid}
\end{figure}

Near the critical point, the extrapolated (to thermodynamic limit) neutral gap in the IPM$_0$ scales linearly as $\Delta_n\propto |J_a-J_a^{\rm critical}|$ as shown in Fig.~\ref{fig:bocdw_gaps}(d) such that the correlation length exponent $\nu=1$, as expected for a 2D Ising critical point. The Ising transition can also be captured by utilizing the fidelity of the ground state defined as 
\begin{equation}
    F=|\langle\psi(J_a)|\psi(J_a+\delta J_a)\rangle|,
\end{equation}
where  $|\psi(J_a)\rangle$ and  $|\psi(J_a+\delta J_a)\rangle$ are the ground state wave functions of the system for the interaction strengths $J_a$ and $J_a+\delta J_a$, respectively, for a small $\delta J_a$~\cite{GU_2010, PhysRevA.98.023615, PhysRevA.102.043710, PhysRevB.89.024424, PhysRevA.87.033609}.

In Fig.~\ref{fig:fid}(a) we plot the fidelity susceptibility defined by the formula
\begin{align}
    \label{fid}
    \chi(J_a) = \lim_{\delta J_a\to 0}\frac{-2\ln F}{(\delta J_a)^2}
\end{align}
as a function of $J_a$ for $J_b=1.75$, and different system sizes such as $L=100,~200,~300$ and $400$. As expected, $\chi/L$ develops a peak for all the system sizes and the peak gradually becomes sharper and shifts towards the critical point (dashed vertical line in Fig.~\ref{fig:fid}(a)) with the increase in system size. This feature can be clearly seen as the divergence of the peak height or the maximum value of $\chi/L$, as a function of system size as shown in the inset of Fig.~\ref{fig:fid}(a). We obtain the critical point by monitoring the shift of the position of the peak $J_{a}$ with increase in system size as shown in  Fig.~\ref{fig:fid}(b). Using a quadratic fitting we extract the extrapolated value of the critical point for $L=\infty$ to be $J_a\sim1.5435$, which is close to the critical point $J_a\sim1.5441$ obtained using the finite size scaling of the structure factor (see Ref.~\cite{PhysRevB.106.L201106}). 

\subsection{\label{sec:dmrgbocdw2} IPM\texorpdfstring{$_\pi$}{} to IN phase transition}

 \begin{figure}
\centering
\includegraphics[width=1\columnwidth]{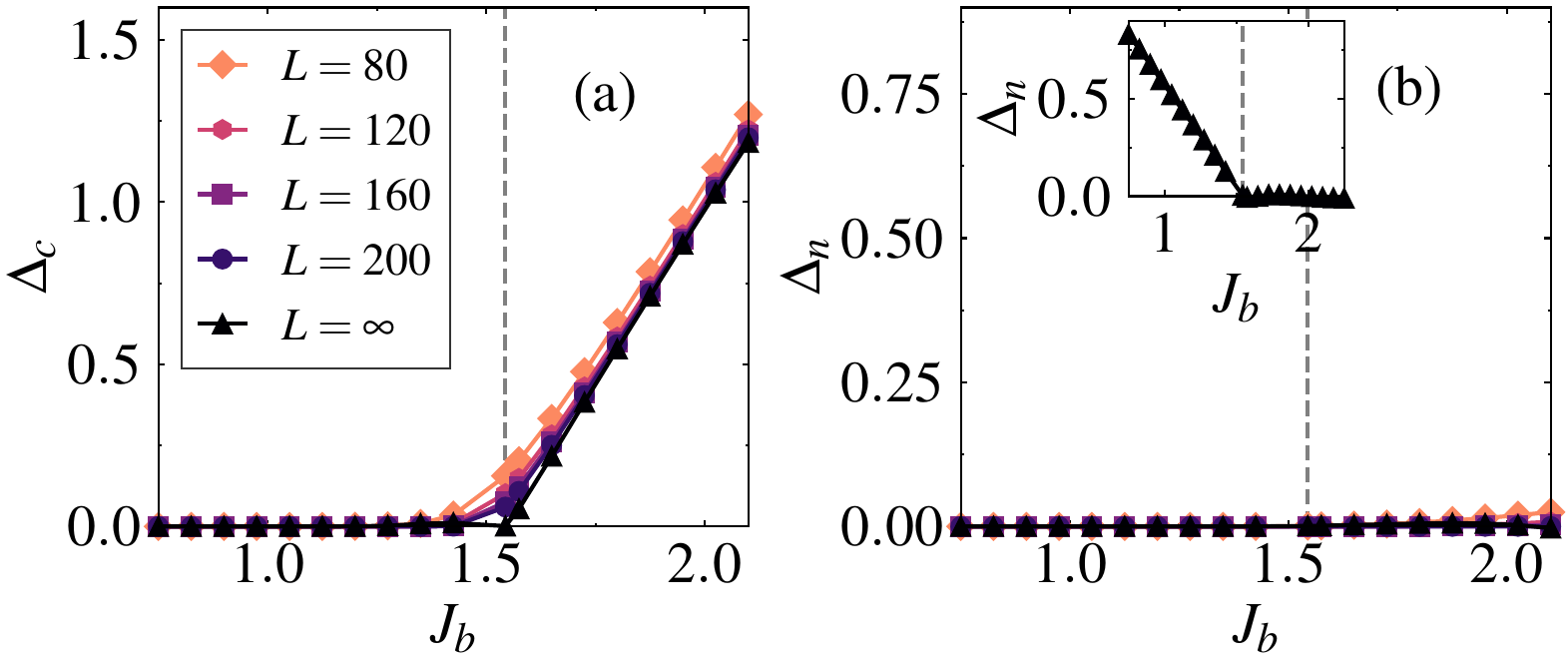}
\caption{(a) Charge gap $\Delta_c$ and (b) neutral gap $\Delta_n$ as a function of $J_b$ for $J_a=1.75$ and varying system sizes. The inset of (b) shows $\Delta_n$ for the system under PBC at $L=\infty$ by extrapolating the data with $L=40, 60, 80$ and $100$. The dashed grey lines mark the critical point corresponding to the IPM$_{\pi}$-IN phase transition, \textit{i.e.,} $J_b=1.5441$.}
\label{fig:ising_gaps2}
\end{figure}

We now investigate the behavior of the charge and neutral gaps across the IPM$_{\pi}$-IN phase transition, which is similar to the IPM$_{0}$-IN transition discussed earlier. As shown in Ref.~\cite{PhysRevB.106.L201106}, both the IPM$_0$ and IPM$_{\pi}$ phases possess identical bulk properties. However, in the IPM$_{\pi}$ phase, the system exhibits gapless edge states, which is absent in the IPM$_{0}$ phase. Here, we consider a cut in the phase diagram of Fig.~\ref{fig:pd_cc} at $J_a=1.75$ and vary $J_b$, for which the critical point of IPM$_{\pi}$-IN phase transition lies at $J_b\sim1.5441$. In Fig.~\ref{fig:ising_gaps2}(a) and (b), we plot the charge gap $\Delta_c$ and neutral gap $\Delta_n$, respectively, as a function of $J_b$ for different system sizes ($L=80,~120,~160,~200$) as well as the finite size extrapolated values at $L=\infty$ to capture this phase transition. As anticipated, due to the presence of gapless edge excitation, $\Delta_c$ vanishes in the IPM$_{\pi}$ regime and there occurs a transition to the IN phase after $J_b\sim1.5441$ where $\Delta_c$ becomes finite. This indicates that inside the IN phase, the zero-energy edge states become energetic and behave just like the bulk states. On the other hand, $\Delta_n$ vanishes throughout for all lengths as a function of $J_b$ as shown in Fig.~\ref{fig:ising_gaps2}(b). This is because both the IPM$_{\pi}$ and IN phases possess degenerate ground states. Thus, the critical point cannot be captured through the behavior of $\Delta_n$. However, under PBC, the degeneracy of the IPM$_\pi$ phase gets lifted and $\Delta_n$ becomes finite. At the same time the ground state of the IN phase still remains degenerate even under PBC. In the inset of Fig.~\ref{fig:ising_gaps2}(b) we plot $\Delta_n$ as a function of $J_b$ in the thermodynamic limit for systems with PBC which show that the gap falls linearly when in the IPM$_\pi$ phase and closes at the critical point, marking an Ising-type transition to the IN phase.

\section{Understanding phase transitions via bosonization}
\label{sec_bosonize}

Having understood the different phases and nature of the phase transitions, we now turn to derive the low-energy critical theories describing the latter. To this end, we follow standard bosonization methods to introduce the soft fermionic modes expanded about the two Fermi points (details discussed in Appendix~\ref{appB}) as shown in Fig.~\ref{fig:low_energy}. The continuum Hamiltonian for $J_a \neq J_b$ case is given by
\begin{align}
    \mathcal{H}=\mathcal{H}_0+\mathcal{H}_{int},
    \label{eqn:continuum_H_full}
\end{align}
where
\begin{align}
    \mathcal{H}_0=-iv_F\int dx\left[ \psi^\dag_R(x)\partial_x \psi_R(x)- \psi^\dag_L(x)\partial_x \psi_L(x) \right]
    \label{eq_freedirac}
\end{align}
is the free Hamiltonian with $\psi_L(\psi_R)$ being the left (right) chiral fermion, $v_F=\lim_{a\rightarrow 0}(J_\perp a)$ is the Fermi velocity ($a$ being the lattice length-scale) and $\mathcal{H}_{int}$ refers to various four-fermion interactions. In particular, it is made up of two terms,
\begin{align}
    \mathcal{H}_{int}=\mathcal{H}_{1}+\mathcal{H}_2,
\end{align}
where
\begin{equation}
    \begin{split}
            \mathcal{H}_1&=\Tilde{\lambda}\int  dx \big[\left(\rho_R  + \rho_L\right)(x)\left(\rho_R  + \rho_L\right)(x+a) \Big]\\
    &-\Tilde{\lambda}\int  dx\Big[ \psi^\dag_L(x)\psi_R(x) \psi^\dag_R(x+a)\psi_L(x+a) + \text{H.c.} \Big]\\
    & - \Tilde{\lambda}\int  dx \Big[ \psi^\dag_L(x)\psi_R(x) \psi^\dag_L(x+a)\psi_R(x+a) + \text{H.c.} \big]
    \end{split}
\end{equation}
with $\Tilde{\lambda}=\lim_{a\rightarrow 0}[(J_b+J_a)a/2]$, and
\begin{align}
    \mathcal{H}_2=&\frac{\Delta}{i\pi a}\int dx\left(:\psi^\dag_R(x)\psi_L(x+a): + \text{H.c.} \right)\nonumber\\
   &- \frac{\Delta}{i\pi a}\int dx\left(:\psi^\dag_L(x)\psi_R(x+a): + \text{H.c.} \right)\nonumber\\
    &- \Delta \int dx :\left[ \left(\rho_R + \rho_L \right)(x) \left(\psi^\dag_R\psi_L + \text{H.c.} \right)(x+a) \right]:\nonumber\\
    &+ \Delta \int dx :\left[\left(\rho_R + \rho_L \right)(x+a) \left(\psi^\dag_R\psi_L + \text{H.c.} \right)(x)  \right]:
    \label{eq_deltasoft}
\end{align}
where $\Delta=\lim_{a\rightarrow 0} a(J_a - J_b)/2$ and $:\mathcal{O} :$ denotes normal ordering of $\mathcal{O}$. Note that $\mathcal{H}_2$ is zero when $J_a=J_b$ whence the theory reduces to the well-known low-energy theory for the XXZ model~\cite{sachdev_2011, Giamarchi_book}. 

The interacting Hamiltonian can now be bosonized~\cite{sachdev_2011,Giamarchi_book}, which is then amenable to systematic renormalization group (RG) treatment leading to the critical theories and the RG flow diagram for the Infrared (IR) fixed points -- both bulk and critical. This is done by introducing bosonized fields, $\Phi(x)$ and $\Theta(x)$~\cite{Giamarchi_book}
\begin{align}
    &\rho_R(x) + \rho_L(x) = - \frac{1}{\pi} ~\nabla \Phi(x), \\
    &\psi_r (x) = \frac{1}{\sqrt{2\pi a}} ~e^{irk_Fx}e^{-i(r\Phi(x)-\Theta(x))}\label{eq:Giamrchi_Bosonization},
\end{align} 
that obey standard commutation relation (see Appendix~\ref{App:E}) and $r=1(-1)$ for $R(L)$. We then get the bosonized form of the continuum Hamiltonian (Eq.~\ref{eqn:continuum_H_full}) as
\begin{equation}\label{eqn:total_bose_Ht}
   \begin{split}
       \mathcal{H}= &\int dx \Bigg[\frac{u}{2\pi}\Big(K(\pi\Pi_\Phi(x))^2 +\frac{1}{K}(\nabla\Phi)^2\Big) \\ 
       & ~~~~~~~~~ + \mathcal{B}^2\Delta\sin (2\Phi) -\frac{\mathcal{B}^2}{2} \Tilde{\lambda} \cos (4\Phi) \Big],
   \end{split}
\end{equation}

where $\mathcal{B} = 1/(\pi a)$ is the bosonization constant and the Luttinger parameter $K$ and the velocity $u$ can be determined from the
Bethe ansatz solution of the XXZ spin chain \cite{Giamarchi_book} in the TLL in terms of the underlying parameters   

\begin{align}\label{eq:total_Bose_Ham_paramt}
    K=\frac{\pi}{2\arccos{(-(J_a+J_b)/2J_\perp})}, \\
    u=\frac{K}{2K-1}\sin\left({\frac{\pi}{2K}}\right) J_\perp.
\end{align}

The action corresponding to the Hamiltonian in Eq.~\ref{eqn:total_bose_Ht} in $(1+1)$-D space is given by:
\begin{equation}\label{eqn:eq3.32}
   \begin{split}
       S= &\int dx d\tau \Big[\frac{1}{2\pi K u}\left((\partial_\tau \Phi)^2+u^2(\nabla\Phi)^2\right) \\
       & ~~~~~~~~~~~ + \mathcal{B}^2\Delta \sin (2\Phi) 
   -\frac{\mathcal{B}^2}{2}\Tilde{\lambda}\cos (4\Phi) \Big],
   \end{split}
\end{equation}
which is precisely the {\it double-frequency sine-Gordon} model studied by G. Delfino, G. Mussardo~\cite{DELFINO1998675} and others~\cite{Fei_2003, Orignac_Giamarchi, fabrizio2000critical}. This model has an Ising transition (with $c=\frac{1}{2}$) for $\Delta\neq 0$ between two gapped phases-- one symmetric and another symmetry broken-- that translates into the transitions between the IPM and the IN phases in Fig.~\ref{fig:pd} which is expected to belong to the Ising universality class according to the strong coupling expansion in the anisotropic limit as explained in Sec.~\ref{sec_IPM}. This is in addition to the $c=1$ transition between the TLL to the IN for $\Delta=0$. This suggests for the possibility that the $c=1$ critical point becomes unstable to $c=\frac{1}{2}$ lines in conformity with the numerical calculations. 

To understand this, we analyse the RG flow equations near the $c=1$ critical point. The one loop $\beta$-functions~\cite{DELFINO1998675, Fei_2003, Orignac_Giamarchi} (see particularly Ref.~\cite{Fei_2003} and also summarized in Appendix~\ref{App:E} for completeness) for $\Delta$, $\tilde\lambda$ and $K$ are given by 

\begin{equation}\label{eqn:total_betat}
   \begin{split}
       &\frac{d\Delta_N}{dl}=(2-K)\Delta_N -c_N \Tilde{\lambda}_N \Delta_N,\\
       &\frac{d\Tilde{\lambda}_N}{dl}=(2-4K)\Tilde{\lambda}_N -c_N \Delta^2_N,\\
       &\frac{dK}{dl} = -a_N(\Delta_N^2 + 4\Tilde{\lambda}_N^2),
   \end{split}
\end{equation}
where
\begin{align}\label{eq:ln_dn}
    \Tilde{\lambda}_N = \frac{\Tilde{\lambda}}{u (\Lambda a)^2},  \ \ \ \ \Delta_N = \frac{\Delta}{u (\Lambda a)^2}
\end{align}
are the re-scaled couplings and 
\begin{align}
    c_N = \frac{4K}{\pi},~~~a_N = \frac{32K^3}{\pi^2}.
\end{align}
Note that both $c_N, a_N >0 $ in the regime of our interest where both $J_a, J_b>0$ (such that $0 < K \leq 1$ in accordance with Eq.~\ref{eq:total_Bose_Ham_paramt}). We focus on this regime unless stated otherwise.

\begin{figure}
    \centering
\includegraphics[width=1\columnwidth]{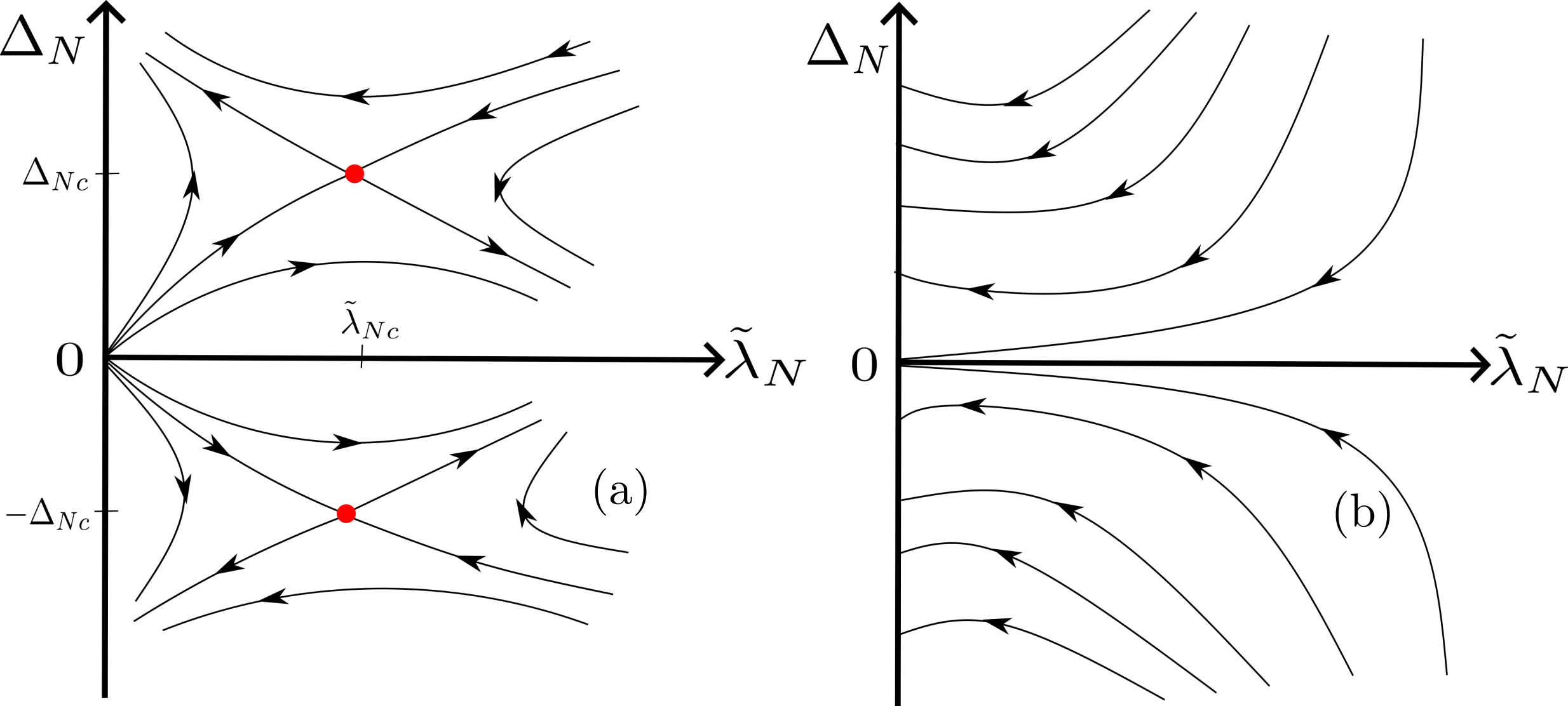}%
\caption{Schematics of RG trajectories illustrating the flow in $\tilde \lambda_N- \Delta_N$ plane for (a) $K < 1/2$  and (b) $K > 1/2$.}\label{fig:RG_flowt}
\end{figure}

The RG-flow diagram in $\Delta-\Tilde{\lambda}$ plane for $K>1/2$ and $K<1/2$ is shown in Fig.~\ref{fig:RG_flowt}. The phase diagram (Fig.~\ref{fig:pd}) of the XXZ chain given by the Hamiltonian in Eq.~\ref{eqn:eq2.1} can be comprehensively understood in terms of the flow diagram via the soft mode fermionic (Eq.~\ref{eqn:continuum_H_full}) and bosonized (Eq.~\ref{eqn:total_bose_Ht}) Hamiltonians.

From Eq.~\ref{eqn:total_betat}, it is clear that  the undimerized $(J_a=J_b)$ line corresponding to $\Delta=0$ corresponds to the known results of the 
spin-$\frac{1}{2}$ XXZ spin-chain whence $K=1/2$ corresponds to the unstable critical point across which the $S^z_i S^z_j$ (or equivalently the $\cos (4\Phi)$) interactions are $\tilde\lambda$ become relevant (for $K<1/2)$. For $K>1/2$, these interactions are irrelevant and we have the gapless TLL in general including the free fixed point corresponding to $K=1$ ({\it i.e.},  $J_a=J_b=0$). On the other hand, for $K<1/2$, the $S^z_iS^z_j$ interactions are relevant and leads to the gapped IN order (Fig.~\ref{fig:pd}). This is obtained by minimizing the bosonized Hamiltonian (Eq.~\ref{eqn:total_bose_Ht}) in the limit of $\tilde\lambda\rightarrow\infty$ whence $\Phi=n\pi/2~(n \in \mathbb{Z})$, so that the IN order parameter $\sim \langle \cos (2\Phi) \rangle = (-1)^n$ (Eq.~\ref{eq:nm_bose}). Thus there are only two inequivalent ground states, for the odd and even values of $n$ (let us take $n=0,1$) such that (see Table~\ref{tab:bose_field_symm}) the TR and translation are spontaneously broken as expected in the IN phase. Hence $K=1/2$ and $\Delta=0$ is the critical point between the TLL described by a $c=1$ CFT and the IN phase. 

Finite dimerization, {\it i.e.,} $\Delta\sim (J_a-J_b)$, as Eq.~\ref{eqn:total_betat} shows, is relevant for the entire regime $(0<K\leq 1)$. This is easy to understand from the first two terms of Eq.~\ref{eq_deltasoft}. Such a fermion bilinear always gaps out the gapless Dirac fermion $(\psi_R,\psi_L)$ described by the free theory in Eq.~\ref{eq_freedirac}. In the bosonized language, this translates into the relevance of $\sin (2\Phi)$ throughout the domain of our interest such that the energy is minimized. {In this case $\Delta$ part of the Hamiltonian is minimized for $\Delta > 0$ whence $\Phi = (4n-1)\pi/4$ ($n\in \mathbb{Z}$), so IPM order parameter $\sim \langle \sin (2\Phi) \rangle = -1~ \forall n$ (Eq.~\ref{Eq:mo_me_bose}) and if $\Delta<0$, $\Phi = (4n+1)\pi/4$ ($n\in \mathbb{Z}$) then $\langle \sin (2\Phi) \rangle = 1~ \forall n$. Thus, TR is not spontaneously broken in this case as is expected for the IPMs. (Incidentally, for $K>2$ which is not within our domain of relevance, both $\Delta$ and $\tilde\lambda$ are irrelevant giving rise to a gapless phase continuously connected to the TLL~\cite{PhysRevB.108.245135}). 

The presence of both the $\sin (2\Phi)$ and $\cos (4\Phi)$ terms lead to interesting interplay with qualitatively and quantitatively different outcomes in the two windows discussed above for the Luttinger parameter, {\it i.e.}, $0<K<1/2$ and $1/2<K<1$ as captured by the above $\beta$-functions captures this in the vicinity of the $c=1$ critical point obtained for 
\begin{align}
    K=1/2,~~~\Delta=\tilde\lambda=0.
\end{align}
Using $K=1/2-\epsilon$ and expanding Eqs.~\ref{eqn:total_betat} about this fixed point to linear orders in the deviation from the fixed point, the $\beta$-functions become
\begin{align}
    \frac{d\Delta_N}{dl}=\Big( \frac{3}{2} + \epsilon \Big) \Delta_N,~~\frac{d\Tilde{\lambda}_N}{dl}=4\epsilon \Tilde{\lambda}_N,~~\frac{d\epsilon}{dl} =0.
    \label{eqn:total_beta_epsilon}
\end{align}

Note that the deviation of the Luttinger parameter, $\epsilon$, does not flow at this order (see below) which makes the analysis near the $c=1$ point simple. With this, combining the first two equations of Eq.~\ref{eqn:total_beta_epsilon} we get:

\begin{equation}\label{eq:rel_delta_lambda}
    \Delta_N \sim \Big( \Tilde{\lambda}_N \Big)^{(\frac{3}{8\epsilon} + \frac{1}{4})}.
\end{equation}

\begin{figure}
    \centering
\includegraphics[width=1\columnwidth]{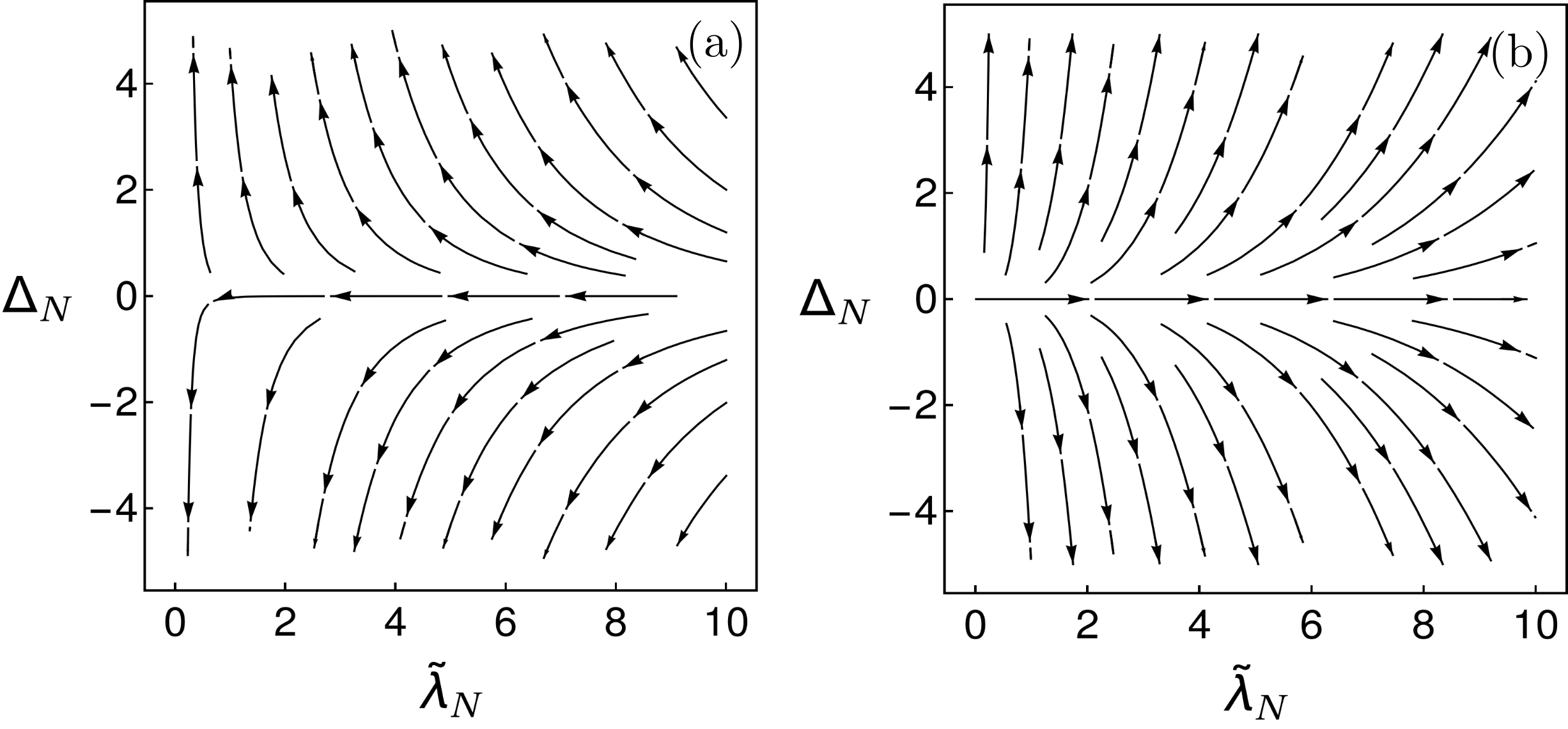}%
\caption{RG flow diagram for (a) $\epsilon=-0.1$  and (b) $\epsilon=+0.1$.}\label{fig:RG_flow_epsilon}
\end{figure}

From the above equation, it is clear that RG flow will be singular at $\epsilon=0$ and is shown in Fig.~\ref{fig:RG_flow_epsilon} for $\epsilon = \pm 0.1$. Clearly for $\epsilon<0$ ({\it i.e. $K>1/2$}) (left panel), the perturbations in $\Delta$ is relevant while $\tilde\lambda$ is irrelevant. Therefore for $K>1/2$, $\tilde\lambda$ flows to zero while $\Delta$ flows to a large value for $J_a\neq J_b$. This then leads to the IPM phases depending on the sign of $\Delta \sim (J_a-J_b) \neq 0$ as explained above-- consistent with the mean-field theory and the 
low-energy Hamiltonian. However, the present bosonization model is incapable of distinguishing between the SPT and the trivial paramagnets. The only exception is the $\Delta=0$ line which is the TLL. 

On the other hand, for $\epsilon>0$ ({\it i.e.}, $K<1/2$) both $\Delta$ and $\tilde\lambda$ are relevant and and they flow to large values as shown  in Fig.~\ref{fig:RG_flow_epsilon} (right panel) leading to the IN. However, for small $\tilde\lambda_N$, {\it i.e.}, near the $c=1$ critical point, the flow is almost horizontal as is also evident from Eq.~\ref{eq:rel_delta_lambda} till $\tilde\lambda_N\sim 1$. This is also evident from the first two Eq.~\ref{eqn:total_betat} which, in the regime $0<K<1/2$, has two new fixed points at ($\tilde{\lambda}_{Nc}, \pm \Delta_{Nc}$) with 
\begin{align}
    \label{eq:fix_pt1} \tilde{\lambda}_{Nc} = \frac{1}{c_N}(2-K),~~~\Delta_{Nc} = \frac{1}{c_N}\sqrt{(2-4K)(2-K)},
\end{align}
as indicated by the red points in the left panel of Fig.~\ref{fig:RG_flowt}. The $c=1$ fixed point is unstable to these new fixed points as shown by the flow. Further, the linear stability analysis about these new fixed points indicate one relevant and one irrelevant direction as indicated in Fig.~\ref{fig:RG_flowt} (left panel). Taking $K\rightarrow 1/2-0^+$, we get $\Delta_{Nc}\rightarrow 0$ and $\tilde\lambda_{Nc}\rightarrow 3\pi/4$ which is not the Gaussian fixed point. While the above values need to be taken carefully and self-consistently accounting for with the flow of $K$, the fact that $\tilde\lambda_{Nc}\neq 0$ for the limit of $\Delta\rightarrow 0^+$ is possibly the reason for the very narrow opening angle of the bifurcation of the $c=1$ critical point to the two $c=\frac{1}{2}$ lines as seen in numerics (see Figs.~\ref{fig:pd} and~\ref{fig:pd_cc}). {Since $\Tilde{\lambda}_N$ denotes the $J_a=J_b$ line in $J_b-J_a$ plane of Fig.~\ref{fig:pd} and $\Delta_N$ is line perpendicular to it, the above analysis based on Eq.~\ref{eq:rel_delta_lambda} explains the narrow angle opening seen in Fig.~\ref{fig:pd}.} This then concludes the bosonization description of the phase diagram and completes our analysis of the low-energy physics of the Hamiltonian in Eq.~\ref{eqn:eq2.1}.

\section{Summary and Outlook}
\label{sec_summary}

We have performed a detailed analysis to understand the phases and phase transitions of a spin-$\frac{1}{2}$ XXZ model with dimerized longitudinal interaction. Numerical studies based on the DMRG method have already revealed the appearance of topological (protected by time-reversal symmetry) and trivial IPM phases and the transition between them through a gap-closing point as a ratio of the dimerization strengths. The topological transition boundary later bifurcates to two such that a symmetry broken IN phase is sandwiched between the two IPM phases in the limit of strong longitudinal interactions~\cite{PhysRevB.106.L201106} leading to the phase diagram shown in Fig. \ref{fig:pd}. In this study, through DMRG calculations we systematically identified the criticalities associated with these phase transitions from the analysis of the central charge, neutral and charge gaps and the fidelity susceptibility measure. Along the undimerized ($J_a=J_b$) line  our calculations reproduce the well known critical point separating the TLL and the symmetry broken (IN) phase via a critical point described by $c=1$ CFT. A remarkable aspect, however, of our findings is the splitting up of the c=1 critical point into two $c=\frac{1}{2}$ Ising critical lines between the topological/trivial paramagnet (IPM) to the symmetry broken (IN) phases.

A mean-field treatment of an equivalent fermionic Hamiltonian, obtained via a Jordan-Wigner transformation, captures the basic structure of the phase diagram where the IPMs maps to trivial and topological insulating phases of an effective SSH model while the IN maps to a charge-density-wave insulator. However, not surprisingly, the mean-field treatment fails to faithfully capture the nature of the phase transitions which then is captured via a renormalization group analysis of the effective low-energy continuum field theory obtained from  bosonization of the spin-Hamiltonian. The resultant bosonized continuum form is given by the so-called {\it double-frequency sine-Gordon} field theory~\cite{DELFINO1998675} which captures the bifurcation of the $c=1$ critical point into two Ising lines, including the scaling of the excitation gap as well as the transition line near the $c=1$ point. Further, we have shown that the appearance of the IPM phases can be explained from the limit of very strong longitudinal interactions and the SPT phase transition between them can be understood from the decorated domain wall picture which becomes most apparent via KW duality to cluster Ising models and deformed quantum AT models. 

Our findings provide the impetus for further exploration of interaction-driven topological phase transition and non-trivial critical phenomena in related interacting models such as the one considered in Ref.~\cite{parida2024}. One can also seek to understand the effect of next-nearest-neighbour couplings as such terms are known to stabilize dimerized phases~\cite{mishra_paramekanti,mishra_rigol,PhysRevB.103.195134}. As the model under consideration can be mapped to the hardcore bosonic Hubbard model, these findings can be tested using state-of-the art experiments involving Rydberg gases~\cite{browayes}. On the other hand, one can also consider set ups with dipolar quantum gases in optical lattices~\cite{ferlaino} to simulate these findings. It is also important to know the fate of these phases and phase transition by relaxing the hardcore constraint - a scenario which may exhibit a much richer phase diagram in this context. 

We end by noting that while our lattice level KW duality shows that the present dimerized XXZ model is dual to the deformed QAT model with both models resulting in the same low-energy continuum double frequency sine-Gordon field theories, the implementation of the symmetries in the dual representations are rather non-trivial~\cite{PhysRevB.108.125135}-- including their effect in the 
low-energy field theories-- and remain to be systematically understood.

\acknowledgements
The authors wish to thank A. Chaubey, H. R. Krishnamurthy and S. Roy for useful discussions and A. Agarwala and S. Santra for a previous collaboration on an associated topic. S.B. acknowledges funding  by Swarna Jayanti fellowship of SERB-DST (India) Grant No. SB/SJF/2021-22/12; DST, Government of India (Nano mission), under Project No. DST/NM/TUE/QM-10/2019 (C)/7; Max Planck Partner group Grant at ICTS. H.N. and S.B. acknowledge the support of the Department of Atomic Energy, Government of India, under project no. RTI4001. T.M. acknowledges support from Science and Engineering Research Board (SERB), Govt. of India, through project No. MTR/2022/000382 and STR/2022/000023. D.S. thanks SERB, India for funding through Project No. JBR/2020/000043. 

\appendix
\section{Fermionic and bosonic representations of the spin-$\frac{1}{2}$ XXZ chain}

In one dimension, the XXZ spin-$\frac{1}{2}$ Hamiltonian (Eq.~\ref{eqn:eq2.1}) can be mapped to either a system of hard-core bosons (used for DMRG) or spinless fermions (used as a starting point for the mean-field theories and the bosonization methods).

\subsection{Spin-$\frac{1}{2}$-(hard-core) boson mapping}

\label{appen_spin-boson}

The representation of the spin-$1/2$ operators in terms of hard-core bosonic creation, annihilation and number operators $a^{\dagger}, a~\text{and}~n=a^\dagger a$~\cite{PhysRev.58.1098}, respectively, is 
\begin{align}
    S^z=\frac{1}{2}-n, ~~~ S^-=a^\dagger,~~~S^+=a.
\end{align}

Using this, the bosonic representation of the spin Hamiltonian (Eq.~\ref{eqn:eq2.1}) is obtained as
\begin{eqnarray}
H=&-& \frac{J_\perp}{2} \sum_{i} (a_{i}^{\dagger}a_{i+1} + \text{H.c.}) \nonumber\\
&+& J_b \sum_{i\in\text{odd}} \left(n_{i}-\frac{1}{2}\right)\left(n_{i+1}-\frac{1}{2}\right) \nonumber\\
&+& J_a \sum_{i\in\text{even}} \left(n_{i}-\frac{1}{2}\right)\left(n_{i+1}-\frac{1}{2}\right).
\label{eq:ham_v1v2}
\end{eqnarray}
The DMRG was done on the above hard-core boson Hamiltonian in Ref. \cite{PhysRevB.106.L201106} as well as this work.

\subsection{Spin-$\frac{1}{2}$-spinless fermion mapping}
\label{appen_spin-fermion}

Using the standard Jordan Wigner transformations~\cite{Jordan1928}:
\begin{equation}\label{eqn:eq5}
\begin{split}
&S_i^z=\frac{1}{2}-c_i^\dag c_i, \\
 &S_i^+=\frac{1}{2}K_i c_i,\\
 &S_i^-=\frac{1}{2}c_i^\dag K_i,
\end{split}
\end{equation}
where $c_i$ are fermions and 
\begin{equation}\label{eqn:eq4}
    K_i=\exp\left[i\pi\sum_{j=1}^{i-1}c_j^\dag c_j\right],
\end{equation}
we can re-write the  XXZ Hamiltonian (Eq.~\ref{eqn:eq2.1}) as in Eq.~\ref{eq_fermionxxz}. We are considering a periodic chain of an even number of sites $L$ at half-filling, so if $L/2$ is odd (even), $\mathbf{S}_{L+1} = \mathbf{S}_1 \implies$  $\hat{c}_{L+1} = - (+) \hat{c}_1$. Thus, the end term of Hamiltonian will be:
\begin{equation}
    S_L^xS_{L+1}^x + S_L^y S_{L+1}^y = -\frac{e^{i\pi L/2}}{2} \Big(c_L^\dagger c_1 + c_1^\dagger c_
    L\Big).
\end{equation}

\section{Degenerate perturbation theory for the anisotropic limits}
\label{appen_degenpert}

Here we provide the details leading up to Eq.~\ref{eqn:eq2.15}. For $J_b\rightarrow \infty$, the unperturbed Hamiltonian is given by Eq.~\ref{eqn:eq2.3} while the perturbation consists of the other two terms in Eq.~\ref{eqn:eq2.1}. These two terms with coupling constants $J_a$ and $J_\perp$ do not mix in the leading order of perturbation and have opposite effects.  

$J_a$ connects two neighbouring (odd/blue) bonds (Fig.~\ref{fig:fig3}) and its effect, for any two neighbouring bonds can be obtained by considering four $\sigma^z$ states 
\begin{align}\label{eq:states}
    \{|+_I+_{I+1}\rangle,|+_I-_{I+1}\rangle,|-_I+_{I+1}\rangle,|-_I-_{I+1}\rangle\}.
\end{align}

The first-order effective Hamiltonian for the above two bond spins is explicitly given by
\begin{equation}\label{eqn:eq2.9}
\tilde{H}_{p1}[I,I+1] = -\frac{J_a}{4}\sigma_I^z\sigma_{I+1}^z,
\end{equation}
which can be obtained by writing out the $4\times 4$ matrix explicitly. Thus, it generates {\it ferromagnetic} Ising interactions between the two bond spins, which align the two $\sigma^z_I$s. Translating this for all sites, we get the first term in Eq.~\ref{eqn:eq2.15}. The ground state of the above Hamiltonian is two-fold degenerate and thus the effect of $J_a$ is to lift the massive degeneracy of $H_b$ by breaking the global $Z_2$ symmetry. Remarkably, in terms of the underlying  $S_i$-spins this is a Ising Neel state that was obtained in the numerical calculations of Ref.~\cite{PhysRevB.106.L201106} in the same regime.

Turning to the $J_\perp$ perturbation, we note that unlike the $J_a$ term, this term acts on both the even and the odd bonds. However, at the leading order, the contribution from each odd bond and its action, on each odd bond, of $|\uparrow,\downarrow\rangle\leftrightarrow |\downarrow\uparrow\rangle$ in terms of the bond spin is $|+_I\rangle\leftrightarrow |-_I\rangle$; this leads to the transverse field term in Eq.~\ref{eqn:eq2.15}.

We now extend this calculation to first order in $1/J_b$.
We find that such a term arises from the action of the $J_\perp$ 
term on the even bonds. Acting on the four possible states on sites
$I$ and $I+1$, the $J_\perp$ term acting on the bond joining them
gives the following:
\begin{eqnarray} 
&& -~ J_\perp (S_{2i}^x S_{2i+1}^x ~+~ S_{2i}^y S_{2i+1}^y) ~|\uparrow \downarrow \rangle_I ~|\uparrow \downarrow \rangle_{I+1} 
\nonumber \\ 
&=& - ~\frac{J_\perp}{2} ~|\uparrow \uparrow \rangle_I ~ 
|\downarrow \downarrow \rangle_{I+1}, \nonumber \\
&& -~ J_\perp (S_{2i}^x S_{2i+1}^x ~+~ S_{2i}^y S_{2i+1}^y) ~|\downarrow \uparrow \rangle_I ~|\downarrow \uparrow \rangle_{I+1} 
\nonumber \\
&=& - ~\frac{J_\perp}{2} ~|\downarrow \downarrow \rangle_I ~
|\uparrow \uparrow \rangle_{I+1}, \nonumber \\
&& -~ J_\perp (S_{2i}^x S_{2i+1}^x ~+~ S_{2i}^y S_{2i+1}^y) ~|\uparrow \downarrow \rangle_I ~|\downarrow \uparrow \rangle_{I+1} ~=~ 0,
\nonumber \\
&& -~ J_\perp (S_{2i}^x S_{2i+1}^x ~+~ S_{2i}^y S_{2i+1}^y) ~|\downarrow \uparrow \rangle_I ~|\uparrow \downarrow \rangle_{I+1} 
~=~ 0. \nonumber \\
\label{pert2} \end{eqnarray}
In the first two equations in Eq.~\ref{pert2}, the state on 
the left hand side (\textit{i.e.,} one of the ground states) has an energy equal to
$-J_b/2$, while the state on the right hand side (\textit{i.e.,} an excited
state) has an energy $J_b/2$. Hence, the energy denominator in 
second-order perturbation theory will be given by $-(J_b/2)-(J_b/2) 
= - J_b$. Further, the matrix element to go between the
left and right hand sides is $-J_\perp/2$, so that the
matrix element squared is $J_\perp^2/4$. Finally, Eq.~\ref{pert2}
shows that only the states $| +_I +_{I+1} \rangle$ and
$| -_I -_{I+1} \rangle$ are affected at second order, while the
states $| +_I -_{I+1} \rangle$ and $| -_I +_{I+1} \rangle$ are not 
affected; this can be encoded by an effective Hamiltonian of the
form $(1/2) (\mathbb{I} ~+~ \sigma_I^z \sigma_{I+1}^z)$. Putting in the
numerical factors, we then find that the second-order term in the 
effective Hamiltonian is given by
\begin{equation}
\tilde{H}_{p2}[I,I+1] ~=~ - ~\frac{J_\perp^2}{8J_b} ~(\mathbb{I} ~+~ 
\sigma_I^z\sigma_{I+1}^z). \end{equation}

Thus at second order in perturbation theory, the coefficient of the first term
in Eq.~\ref{eqn:eq2.15} changes from $-J_a/4$ to $-(J_a/4)-
(J_\perp^2/(8J_b))$, while the coefficient of the second term remains unchanged at $-
J_\perp/2$. The critical line of the effective transverse field Ising 
chain is given by the condition that the two coefficients are equal.
This gives
\begin{equation} J_a ~=~ 2 J_\perp ~-~ \frac{J_\perp^2}{2 J_b},
\end{equation}
which is Eq.~\ref{pt2} in the main text.


\section{Duality transformation from dimerized XXZ chain to quantum Ashkin-Teller (QAT) model} \label{app:XXZ_to_QAT}

We can map the dimerized XXZ Hamiltonian (Eq.~\ref{eqn:eq2.1}) to quantum Ashkin-Teller model (QAT) using Kramers-Wannier (KW) duality transformations following Ref.~\cite{PhysRevB.24.5229}. We reproduce some of the details for completeness. Using ${\bf S}_i = \boldsymbol{\mu}_i / 2$, Eq.~\ref{eqn:eq2.1} can be written as 

\begin{equation}\label{eq: H_XXZ_mu}
    \begin{split}
        H = & -\frac{J_\perp}{4}\sum_i [\mu^x_{2i} \mu^x_{2i-1} + \mu^y_{2i} \mu^y_{2i-1}] \\
        & -\frac{J_\perp}{4}\sum_i [\mu^x_{2i} \mu^x_{2i+1} + \mu^y_{2i} \mu^y_{2i+1}] \\
        & + \frac{J_b}{4}\sum_i \mu^z_{2i} \mu^z_{2i-1} + \frac{J_a}{4}\sum_i \mu^z_{2i} \mu^z_{2i+1}.
    \end{split}
\end{equation}

Now, turning to the odd (blue) bonds denoted by $I\equiv (2i-1,2i)$ as done in the main text above and in Eq.~\ref{eqn:eq2.3}, we relabel the sites as (Fig.~\ref{fig:XXZ_QAT})
\begin{align}
    &2i-1 \rightarrow I - 1/4\equiv (I-1)+3/4, ~~~~ 2i\rightarrow I+1/4, \nonumber\\
    &~~~~2i+1 \rightarrow I+3/4,
    \label{eq_xxzatrelabel}
\end{align}  
such that $I$ lies in the middle of all the odd (blue) bonds and $I+\frac{1}{2}$ lies at the middle of even (red) bonds. Note that we could have also used the midpoint of the even (red) bonds to perform the above relabeling. 

Finally performing a unitary rotation by $\pi/2$ about $\mu^x$, such that
\begin{align}\mu^z \rightarrow \mu^y, ~~ \mu^y \rightarrow -\mu^z,~~ \mu^x\rightarrow \mu^x,
\label{eq_xxzqatrot}
\end{align}
we get

\begin{equation}\label{eq: H_XXZ_mu_j}
    \begin{split}
        H = & -\frac{J_\perp}{4}\sum_I \left[\mu^x_{I+\frac{1}{4}} \mu^x_{(I-1)+\frac{3}{4}} + \mu^z_{I+\frac{1}{4}} \mu^z_{(I-1)+\frac{3}{4}}\right] \\
        & -\frac{J_\perp}{4}\sum_I \left[\mu^x_{I+\frac{1}{4}} \mu^x_{I+\frac{3}{4}} + \mu^z_{I+\frac{1}{4}} \mu^z_{I+\frac{3}{4}}\right] \\
        & + \frac{J_b}{4}\sum_I \mu^y_{I+\frac{1}{4}} \mu^y_{(I-1)+\frac{3}{4}} + \frac{J_a}{4}\sum_I \mu^y_{I+\frac{1}{4}} \mu^y_{I+\frac{3}{4}}.
    \end{split}
\end{equation}

\begin{figure}
    \centering
    \includegraphics[width=1\columnwidth]{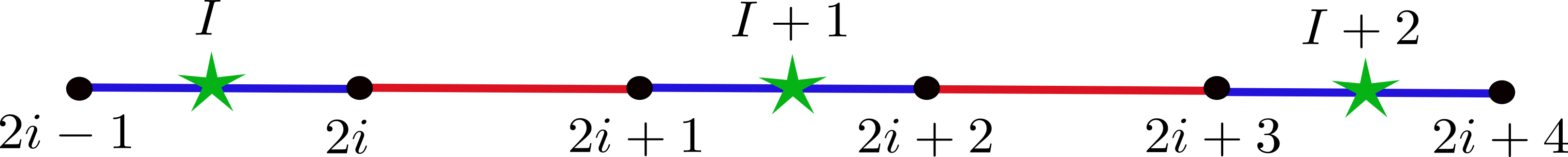}
    \caption{XXZ spins ($\mu_i$) reside on sites $2i-1,~2i,~ 2i+1,\ldots$ while QAT spins ($\sigma_I, ~ \Tilde{\tau}_I$) are placed on sites $I,~I+1,\ldots$ of dual lattice (which are centre of odd (blue) bonds in this case).}
    \label{fig:XXZ_QAT}
\end{figure}

Now we define the following KW duality transform to map $\mu$ spins to $\rho$ spins (defined at sites $I$ and links $I+1/2$ of dual lattice):

\begin{align}\label{eq:KW1}
    &\mu^z_{I+\frac{1}{4}} = \prod_{k=1}^{I} \rho^x_k,~~~~\mu^x_{I+\frac{1}{4}} = \rho^z_{I+\frac{1}{2}} \rho^z_{I},\nonumber\\
    &\mu^y_{I+\frac{1}{4}}= -\rho^y_{I}\rho^z_{I+\frac{1}{2}} \prod_{k=1}^{I - 1/2} \rho^x_{k},
\end{align}
and
\begin{align}\label{eq:KW2}
    &\mu^z_{I+\frac{3}{4}} = \prod_{k=1}^{I+1/2} \rho^x_k,~~~~\mu^x_{I+\frac{3}{4}}=\rho^z_{I+1}\rho^z_{I+\frac{1}{2}},\nonumber\\
    &\mu^y_{I+\frac{3}{4}}= -\rho^y_{I+\frac{1}{2}}\rho^z_{I+1} \prod_{k=1}^{I} \rho^x_{k},
\end{align}
where we consider an open system starting at $I=1$ and the product is over all $\rho^x$ on the sites $I$ and bonds $I+1/2$. Using this transformation, we can 
re-write Eq.~\ref{eq: H_XXZ_mu_j} as 
\begin{equation}\label{eq: H_XXZ_rho_j}
    \begin{split}
        H = & -\frac{J_\perp}{4}\sum_I \left[\rho^z_{I+\frac{1}{2}} \rho^z_{I-\frac{1}{2}} + \rho^x_I\right] \\
        & -\frac{J_\perp}{4}\sum_I \left[\rho^z_I \rho^z_{I+1} + \rho^x_{I+\frac{1}{2}}\right] \\
        & - \frac{J_b}{4}\sum_I \rho^z_{I-\frac{1}{2}} \rho^x_{I} \rho^z_{I+\frac{1}{2}} - \frac{J_a}{4}\sum_I \rho^z_{I} \rho^x_{I+\frac{1}{2}}\rho^z_{I+1}.
    \end{split}
\end{equation}
Let us now re-label the spins on the bonds ($I+1/2,~ I+3/2, \ldots$) by $\tilde\tau^\alpha$ and the ones at the sites  ($I, I+1, \ldots$) by $\sigma^\alpha$ where $\alpha=x,y,z$. Then we can re-write Eq.~\ref{eq: H_XXZ_rho_j} as

\begin{equation}\label{eq: H_XXZ_rho_ST}
    \begin{split}
        H = & -\frac{J_\perp}{4}\sum_I [\Tilde{\tau}^z_{I+\frac{1}{2}} \Tilde{\tau}^z_{I-\frac{1}{2}} + \sigma^x_{I}] \\
        & -\frac{J_\perp}{4}\sum_I \left[\sigma^z_{I} \sigma^z_{I+1} + \Tilde{\tau}^x_{I+\frac{1}{2}}\right] \\
        & - \frac{J_b}{4}\sum_I \Tilde{\tau}^z_{I-\frac{1}{2}} \sigma^x_{I} \Tilde{\tau}^z_{I+\frac{1}{2}} - \frac{J_a}{4}\sum_I \sigma^z_{I} \Tilde{\tau}^x_{I+\frac{1}{2}}\sigma^z_{I+1}.
    \end{split}
\end{equation}

Finally we apply KW duality transformation only on the $\tilde\tau$ spins as

\begin{align}
    \Tilde{\tau}^z_{I+\frac{1}{2}} &= \prod_{k=1}^{ I}\tau^x_{k},\\
    \Tilde{\tau}^x_{I+\frac{1}{2}} &= \tau^z_{I} \tau^z_{I+1},
\end{align}
such that Eq.~\ref{eq: H_XXZ_rho_ST} becomes
\begin{equation}\label{eq: H_QAT}
    \begin{split}
        H = & -\frac{J_\perp}{4}\sum_I \Big[ \tau^x_{I} +  \sigma^x_{I} + \tau^z_{I}\tau^z_{I+1} +  \sigma^z_{I}\sigma^z_{I+1} \Big]\\
        &- \frac{J_b}{4}\sum_I \tau^x_{I} \sigma^x_{I} - \frac{J_a}{4}\sum_I \tau^z_{I} \tau^z_{I+1} \sigma^z_{I} \sigma^z_{I+1},
    \end{split}
\end{equation}
which we re-write as
\begin{align}
    H=&-\frac{J_\perp}{4}\sum_I \Big[ \tau^x_{I} +  \sigma^x_{I}+\frac{J_b}{J_\perp}\tau^x_{I} \sigma^x_{I}\Big]\nonumber\\
    &-\frac{J_\perp}{4}\sum_I\Big[\tau^z_{I}\tau^z_{I+1} +  \sigma^z_{I}\sigma^z_{I+1}+\frac{J_a}{J_\perp}\tau^z_{I} \tau^z_{I+1} \sigma^z_{I} \sigma^z_{I+1} \Big]
    \label{eq_qatm1}
\end{align}
which is a modified QAT model~\cite{PhysRev.64.178,PhysRevB.24.230,PhysRevB.91.165129,PhysRevB.24.5229,PhysRevB.108.184425}. Note that we could have defined the $I$ sites on the even (red) bonds in Fig.~\ref{fig:XXZ_QAT} and would have gotten a different, but related, QAT model.

For our purpose, it is useful to re-write the above Hamiltonian as
\begin{align}
    H=H_{\rm QAT}+H_{\rm pert}
    \label{eq_qatm}
\end{align}
where
\begin{align}
    H_{\rm QAT}=&-\frac{J_\perp}{4}\sum_I \Big[ \tau^x_{I} +  \sigma^x_{I}+\frac{J_a}{J_\perp}\tau^x_{I} \sigma^x_{I}\Big]\nonumber\\
    &-\frac{J_\perp}{4}\sum_I\Big[\tau^z_{I}\tau^z_{I+1} +  \sigma^z_{I}\sigma^z_{I+1}+\frac{J_a}{J_\perp}\tau^z_{I} \tau^z_{I+1} \sigma^z_{I} \sigma^z_{I+1} \Big]
\end{align}
is the critical QAT model which has a transition from a critical phase to a gapped partially ordered phase at $J_a=J_b=J_\perp$. For the entire window $J_a=J_b\in [0,1)$ connecting the two endpoints the critical phase has varying exponents and is described by a CFT with central charge $c=1$ with $J_a=J_b=0$ being the free fermion CFT or the $XY$-model in Eq.~\ref{eqn:eq2.1}. 

Away from the symmetric line, additionally, we have 
\begin{align}
    H_{\rm pert}=-\frac{J_b-J_a}{4}\sum_I\tau^x_I\sigma_I^x.
\end{align}

Starting with Eq.~\ref{eq_qatm}, we can reach the same low-energy theory of double frequency Sine-Gordon model (Eq.~\ref{eqn:eq3.32}) following the methods introduced in Ref.~\cite{fabrizio2000critical}. Instead of doing that, we study the anisotropic limit of the dual Hamiltonian in Eq.~\ref{eq_qatm} to obtain the trivial and topological paramagnets-- IPM$_0$ and IPM$_\pi$ phases-- as well as the associated phase transitions into the IN phase.

\subsection{Anisotropic limit of the modified QAT}

To be precise, let us consider the anisotropic limit by taking $J_b\rightarrow\infty$ such that we re-write Eq.~\ref{eq_qatm1} as 
\begin{align}
    H&=-\frac{J_b}{4}\sum_I\tau^x_I\sigma_I^x\nonumber\\
    &-\frac{J_\perp}{4}\sum_I\left[\tau^x_I+\sigma_I^x+\tau^z_I\tau^z_{I+1}+\sigma^z_I\sigma^z_{I+1}\right]\nonumber\\
    &-\frac{J_a}{4}\sum_I \sigma^z_I\tau^z_I\sigma^z_{I+1}\tau^z_{I+1}.
\end{align}
In the limit of $J_b\rightarrow\infty$, at each site, the ground state manifold is now given by a doublet $|\tau^x=\pm 1,\sigma^x=\pm 1\rangle$. This is similar, but not exactly equal (due to the symmetry implementation), to Eq.~\ref{eq_GSising}. Now it is easy to show (following the methods in Appendix~\ref{appen_degenpert}) that within this manifold, to the leading order in $J_\perp$ and $J_a$, the effective Hamiltonian is given by
\begin{align}
    H_\text{eff}=-\frac{J_a}{4}\sum_I\sigma^z_I\sigma^z_{I+1}-\frac{J_\perp}{2}\sum_I\sigma^x_I
\end{align}
where to avoid further proliferation of notations, we note that the low-energy doublet can be either represented by denoting either $\tau^x=\pm 1$ or $\sigma^x=\pm 1$ and we have used $\sigma$ to denote the above low-energy manifold. 

The above Hamiltonian is an equivalent version of Eq.~\ref{eqn:eq2.15}. Hence it has a similar phase diagram as a function of $J_\perp/J_a$ that describes the IPM$_0$ and the IN phases and the intervening Ising phase transition. We do not explore the equivalences of the phases further, however, since the implementation of the symmetries on the dual Hamiltonian is rather subtle and needs careful understanding due to the non-local KW duality~\cite{PhysRevB.108.125135} which is not the main focus of this work.

\subsection{The duality and decorated domain-wall construction for the SPT realized in \texorpdfstring{IPM$_\pi$}{}}

\begin{figure}
    \centering
    \includegraphics[width=1\columnwidth]{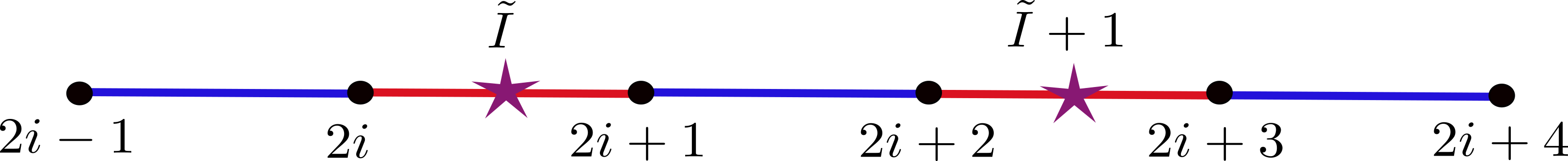}
    \caption{XXZ spins ($\mu_i$) reside on sites $2i-1,~2i,~ 2i+1,\ldots$ while KW transformed spins ($\sigma_{\tilde I}, ~ \tilde \tau_{\tilde I+1/2}$) are placed on sites $\tilde I$ and bonds $\tilde I+1/2$ of dual lattice (sites $\tilde I$ are centre of even (red) bonds in this case).}
    \label{fig:XXZ_QAT2}
\end{figure}

A similar analysis will also lead to an effective Hamiltonian similar to Eq.~\ref{eqn:eq2.19b} that describes the IPM$_\pi$ and IN as well as their intermediate phase transition. However, to understand the topological nature of IPM$_\pi$, we find it easiest to use an alternate, but closely related, version of the XXZ $\leftrightarrow$ QAT duality transform compared to the previous subsection. We briefly outline this below and connect it to the decorated domain-wall picture of Subsection~\ref{sec_ddw}. To this end, we start with Eq.~\ref{eq: H_XXZ_mu} for an open chain as shown in Figs.~\ref{fig:fig3} and~\ref{fig:XXZ_QAT}. Then, instead of following the rest of the duality which leads to a representations of the AT spins on the odd (blue) bonds, we consider a similar version where the AT spins are on the even (red) bonds similar to the discussion around Eq.~\ref{eqn:eq2.7pi}  and Fig.~\ref{fig:top_chain} in the main text. This is obtained by relabelling the sites as 
\begin{align}
    &2i-1 \rightarrow \tilde{I} - 3/4,~~~~ 2i\rightarrow \tilde{I}-1/4,\nonumber\\
    &2i+1 \rightarrow \tilde{I}+1/4\equiv (\tilde{I}+1)-\frac{3}{4}
\end{align} 
instead of Eq.~\ref{eq_xxzatrelabel}. It is clear that this is akin to choosing the bond spins on the even (red) bonds similar to that discussed in Fig.~\ref{fig:top_chain} which is more explicitly shown in Fig.~\ref{fig:XXZ_QAT2}.

Following the above relabelling and skipping the rotation in Eq.~\ref{eq_xxzqatrot}, we use the following KW transformation (which are variants of Eqs.~\ref{eq:KW1} and~\ref{eq:KW2})
\begin{align}\label{eq:KW3}
    &\mu^z_{\tilde I-\frac{1}{4}} = \prod_{k=1}^{\tilde I-1/2} \rho^x_k,~~~~\mu^x_{\tilde I-\frac{1}{4}} = \rho^z_{\tilde I-\frac{1}{2}} \rho^z_{\tilde I},\nonumber\\
    &\mu^y_{\tilde I-\frac{1}{4}}= -\rho^y_{\tilde I-\frac{1}{2}}\rho^z_{\tilde I} \prod_{k=1}^{\tilde I-1} \rho^x_{k},
\end{align}
and
\begin{align}\label{eq:KW4}
    &\mu^z_{\tilde I-\frac{3}{4}} = \prod_{k=1}^{\tilde I-1} \rho^x_k,~~~~\mu^x_{\tilde I-\frac{3}{4}}=\rho^z_{\tilde I-1}\rho^z_{\tilde I-\frac{1}{2}},\nonumber\\
    &\mu^y_{\tilde I-\frac{3}{4}}= -\rho^y_{\tilde I-1}\rho^z_{\tilde I-\frac{1}{2}} \prod_{k=1}^{\tilde I-3/2} \rho^x_{k}
\end{align}
on the Hamiltonian in Eq.~\ref{eq: H_XXZ_mu} to get
\begin{equation}\label{eq: H_XXZ_rho_j2}
    \begin{split}
        H = & -\frac{J_\perp}{2}\sum_{\tilde I} \left[\rho^z_{\tilde I-\frac{1}{2}}\Big(\frac{1-\rho_{\tilde I}^x}{2}  \Big) \rho^z_{\tilde I+\frac{1}{2}} \right] \\
        & -\frac{J_\perp}{2}\sum_{\tilde I} \left[\rho^z_{\tilde I -1} \Big(\frac{1-\rho_{\tilde I-\frac{1}{2}}^x}{2}  \Big) \rho^z_{\tilde I} \right] \\
        & + \frac{J_b}{4}\sum_{\tilde I} \rho^x_{\tilde I-\frac{1}{2}} + \frac{J_a}{4}\sum_{\tilde I} \rho^x_{\tilde I},
    \end{split}
\end{equation}
where we explicitly choose the open system as shown in Fig.~\ref{fig:XXZ_QAT3} such that we consider an open system starting at $\tilde I=1$ and the product is over all $\rho^x$ on the sites $\tilde I$ and bonds $\tilde I+1/2$.
\begin{figure}
    \centering
    \includegraphics[width=1\columnwidth]{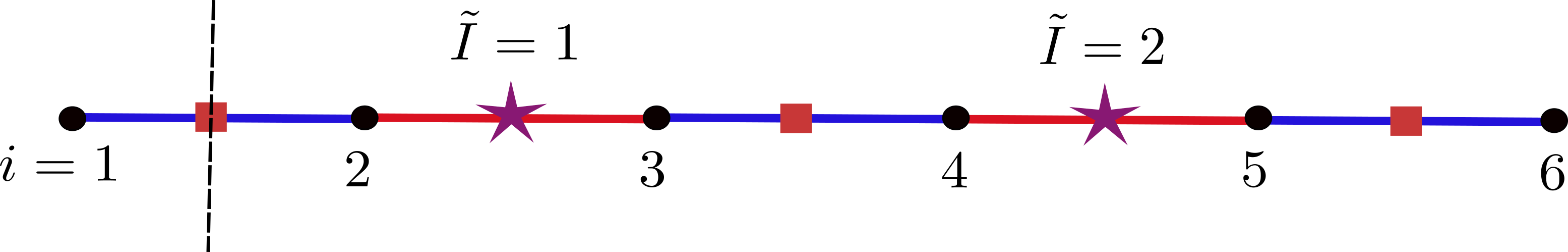}
    \caption{{Spin chain for $J_a\rightarrow \infty$ limit, starting with $i=1$ and $\tilde I =1$. The domain wall at $\tilde \tau^z_{1/2}$ can reach the boundary in this case.}}
    \label{fig:XXZ_QAT3}
\end{figure}

{Let us now re-label the spins on the bonds ($\tilde I+1/2,~ \tilde I+3/2, \ldots$) by $\tilde \tau^\alpha$ and the ones at the sites  ($\tilde I, \tilde I+1, \ldots$) by $ \sigma^\alpha$ where $\alpha=x,y,z$. Then we can re write Eq.~\ref{eq: H_XXZ_rho_j} as
\begin{equation}\label{eq: H_XXZ_rho_j_ST}
    \begin{split}
        H = & -\frac{J_\perp}{2}\sum_{\tilde I} \left[\tilde \tau^z_{\tilde I-\frac{1}{2}}\Big(\frac{1-{\sigma}_{\tilde I}^x}{2}  \Big) \tilde \tau^z_{\tilde I+\frac{1}{2}} \right] \\
        & -\frac{J_\perp}{2}\sum_{\tilde I} \left[{\sigma}^z_{\tilde I -1} \Big(\frac{1-\tilde \tau_{\tilde I-\frac{1}{2}}^x}{2}  \Big) {\sigma}_{\tilde I} \right] \\
        & + \frac{J_b}{4}\sum_{\tilde I} \tilde \tau^x_{\tilde I-\frac{1}{2}} + \frac{J_a}{4}\sum_{\tilde I} {\sigma}^x_{\tilde I}.
    \end{split}
\end{equation}

In the limit of $J_a\rightarrow\infty$, the fourth term in Eq.~\ref{eq: H_XXZ_rho_j_ST} forces  ${\sigma}_{\tilde I}^x=-1~\forall~\tilde I$, thus effective Hamiltonian becomes:
\begin{equation}\label{eff_H_Jb}
    H_{\text{eff}}[J_a\rightarrow\infty]  = -\frac{J_\perp}{2}\sum_{\tilde I} \tilde\tau^z_{\tilde I-\frac{1}{2}} \tilde\tau^z_{\tilde I+\frac{1}{2}} + \frac{J_b}{4} \sum_{\tilde I} \tilde\tau^x_{\tilde I-\frac{1}{2}}.
\end{equation}
This is the dual form of the transverse field Ising model derived earlier. Thus the ordered phase of the above model denotes the IPM.

To understand the nature of the IPM, let us go back to Fig.~\ref{fig:XXZ_QAT3} which starts from $\tilde{I}=1$. Note that since the domain wall, $\tilde\tau^z_{\tilde{I}+1/2}$ has condensed for $J_\perp/J_b>1$ on the odd (blue) bonds, the underlying spins on the even (red) bonds are disordered as expected in the IPM phase. But this basically means that the end-spin-$\frac{1}{2}$ at $i=1$-- which is half of a red bond, is free to fluctuate as the domain wall at $\tilde\tau^z_{1/2}$ can reach the boundary as discussed in the main text. Since the edge spin transforms under time reversal as a Kramers doublet, this gives rise to the two-fold degenerate edge mode as expected in the SPT, {\it i.e.}, IPM$_\pi$.

In the limit of $J_b \rightarrow \infty$, the third term of Eq.~\ref{eq: H_XXZ_rho_j_ST} forces $\tilde\tau_{\tilde{I}-1/2}^x=-1~\forall~\tilde{I}$, thus effective Hamiltonian becomes:

\begin{equation}\label{eff_H_Ja}
    H_{\text{eff}}[J_b\rightarrow\infty]  = -\frac{J_\perp}{2}\sum_{\tilde{I}} {\sigma}^z_{\tilde{I}} {\sigma}^z_{\tilde{I}+1} + \frac{J_a}{4} \sum_{\tilde{I}} {\sigma}^x_{\tilde{I}}.
\end{equation}

Now in the regime $J_\perp/J_b>1$, the domain walls on the even (red) bonds, {\it i.e.} ${\sigma}^z_{\tilde{I}}$, condense. Such that the spins, now sitting on the odd (blue) bonds are disordered. But this immediately means with respect to Fig.~\ref{fig:XXZ_QAT3}, that the domain wall cannot reach the boundary and now the boundary spin is made up of a non-Kramers doublet which does not have a gapless edge mode. This is therefore the topologically trivial phase IPM$_0$.

\section{Details of Numerical Results}
\label{appen_centralcharge}

We perform the numerical simulations by using the density matrix renormalization group (DMRG) method ~\cite{PhysRevLett.69.2863, RevModPhys.77.259} based on matrix product states (MPS) ansatz~\cite{Rommer_1997, Schollw_ck_2011, Catarina2023}, which is a sophisticated numerical technique to efficiently solve low dimensional quantum many-body systems. In order to ensure sufficient accuracy of this method, we run the algorithm with $10$ number of sweeps and a maximum of $500$ bond dimensions. We consider a fixed number of particles at density $\rho = N/L = 1/2$ with systems consisting of up to $L=1000$ lattice sites. Unless otherwise explicitly mentioned, we consider open boundary conditions. 

\begin{figure}
\centering
\includegraphics[width=1\columnwidth]{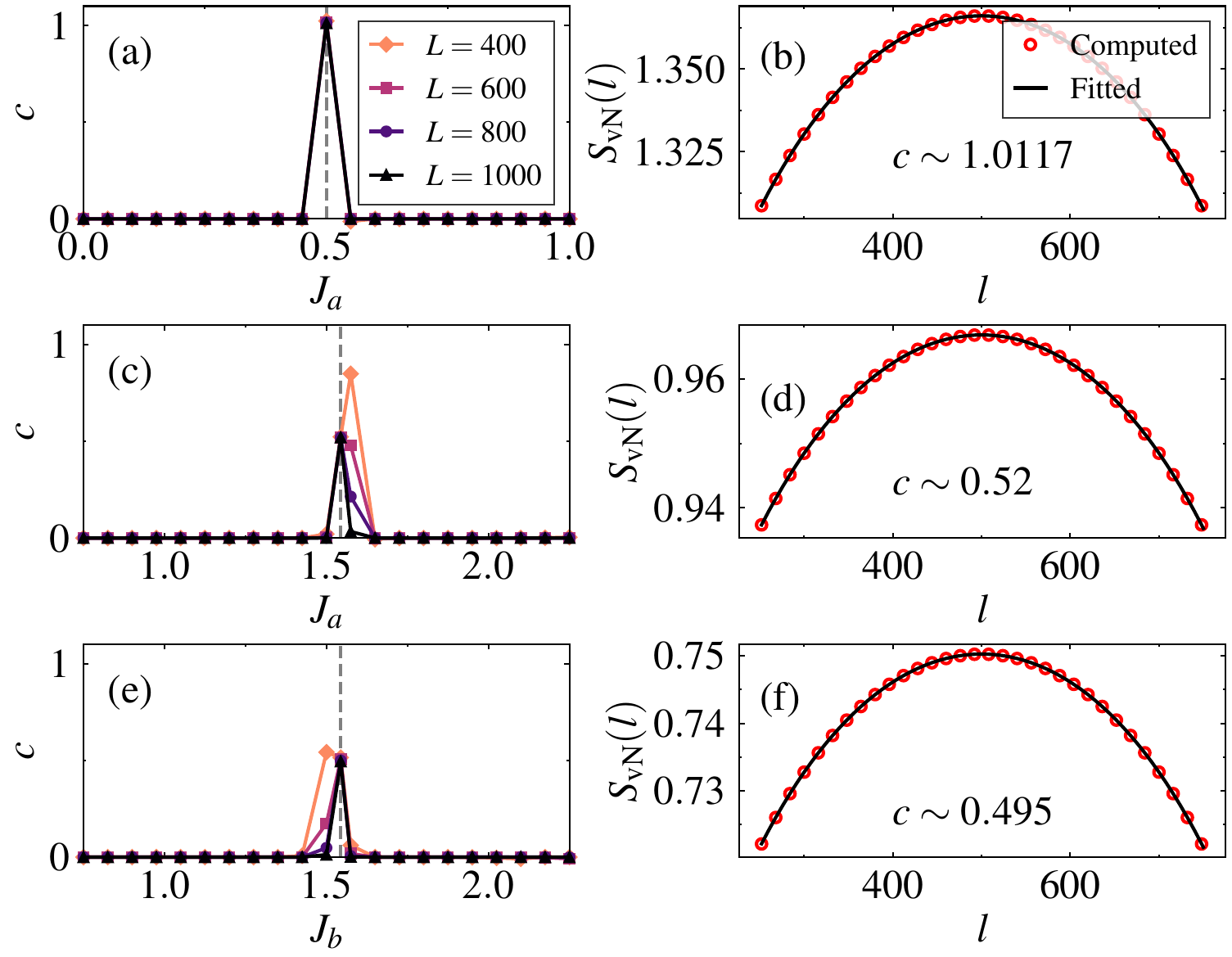}
\caption{Central charge $c$ is plotted (a) as a function of $J_a$ for $J_b=0.5$, (c) as a function of $J_a$ for $J_b=1.75$ and (e) as a function of $J_b$ for $J_a=1.75$ and varying system sizes. The dashed grey lines mark the critical points corresponding to the IPM$_0$-IPM$_{\pi}$, IPM$_0$-IN and IPM$_{\pi}$-IN phase transitions, \textit{i.e.,} $J_a=0.5$, $J_a=1.5441$ and $J_b=1.5441$, respectively. (b), (d) and (f) show $S_{\rm vN}(l)$ plotted as a function of lattice sites $l$ at the three critical points, respectively, for a system size $L=1000$.}
\label{fig:cc_tpt}
\end{figure}

\begin{figure}
\centering
\includegraphics[width=1\columnwidth]{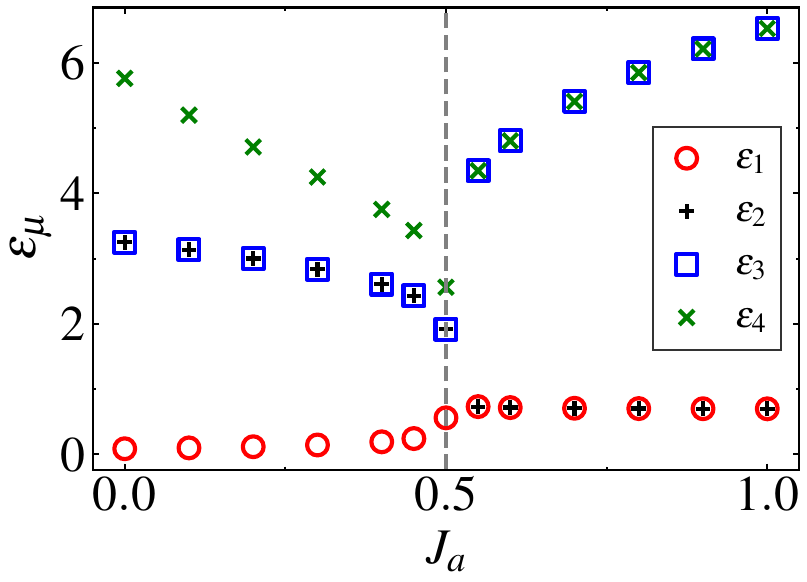}
\caption{The lowest four eigenvalues of the entanglement spectrum $\varepsilon_{\mu}$ as a function of $J_a$ for $J_b=0.5$ on a system of size $L=1000$. The dashed grey line marks the critical point corresponding to the topological phase transition, \textit{i.e.,} $J_a=0.5$.}
\label{fig:entspec}
\end{figure}

First of all we focus on to understand the criticality associated with the  IPM$_0$-IPM$_\pi$ and IPM-IN phase transitions by estimating the central charge $c$ which is defined using the Cardy-Calabrese formula~\cite{Calabrese_2009} 
\begin{align}
\label{eq:ent2}
    S_{\rm vN}(l) = \frac{c}{6}\ln \left[\frac{L}{\pi}\sin\left(\frac{\pi l}{L}\right) \right] + g,
\end{align}
where,  $S_{\rm vN}(l)$ is the von Neumann (vN) entropy, $c$ is the central charge and $g$ is a non-universal constant. The $S_{\rm vN}(l)$ for a subsystem of length $l$ is defined as 
\begin{align}
S_{\rm vN}(l) = -\text{Tr}_l[\rho_l\ln\rho_l],
\label{eq:ent}
\end{align}
where $\rho_l$ is the reduced density matrix of the subsystem $l<L$. Additionally, the entanglement spectrum defined by $\varepsilon_{\mu}=-2\ln \lambda_{\mu}$ where $\lambda_{\mu}$ are the singular values of $\rho_l$, provides information about ground state phases and phase transitions.

The central charge is calculated by an appropriate fitting of $S_{\rm vN}(l)$ with respect to the lattice sites $l$. For our calculations, we consider every alternate bonds and discard the boundary effects by considering $l$ in the range $L/4$ to $3L/4$. Note that this formula works for OBC only.

In Figs.~\ref{fig:cc_tpt}(a), (c) and (e), we plot the values of $c$ for three exemplary cuts across the phase diagram of Fig.~\ref{fig:pd} at $J_b=0.5$, $J_b=1.75$ and $J_a=1.75$ which correspond to the IPM$_0$-IPM$_\pi$, IPM$_0$-IN and IPM$_\pi$-IN phase transitions, respectively. We obtain that at the IPM$_0$-IPM$_\pi$ transition point the central charge takes a value $c\sim1$. However, for both the IPM$_0$-IN and the IPM$_\pi$-IN transition points the central charge is $c \sim 1/2$. As expected $c=0$ inside all the gapped phases. The value of $c$ is obtained by a fitting Eq.~\ref{eq:ent2} to the numerically obtained von Neumann entropy $S_{vN}(l)$ as shown in Figs.~\ref{fig:cc_tpt}(b), (d) and (f) at the three critical points,\textit{ i.e.,} $J_a=0.5$, $J_a=1.5441$ and $J_b=1.5441$, respectively. The exact numerical values of the central charge turn out to be $c\sim1.0117$, $c\sim0.52$ and $c\sim0.495$, respectively, which are close to the expected values from CFT.

The extracted values of $c$ is shown across the different phase boundaries of Fig.~\ref{fig:pd} in Fig.~\ref{fig:pd_cc}. 

This analysis indicates that the $c=1$ critical line slowly bifurcates into two $c=\frac{1}{2}$ 
lines with increase in the interaction strengths. However, close to the bifurcation point, $c$ deviates from its expected values due to finite size effects.

Alternatively, we also take a straightforward route for computing the central charge to minimize error associated to fitting of functions used in the previous approach. To this end we first compute $S_{\text{vN}}(L/2)$ and $S_{\text{vN}}(L/2-2)$ using Eq.~\ref{eq:ent2} and then subtract one from the other to obtain the formula
\begin{align}
\label{eq:ent3}
    c = \frac{6[S_{\rm vN}(L/2-2)-S_{\rm vN}(L/2)]}{\ln\{\cos[\pi/(L/2)]\}} \equiv c^*.
\end{align}
This formula is advantageous compared to the previous one because the non-universal constant $g$ does not appear here, and the boundary effects are discarded by default since we consider only two entanglement entropies in the middle of the lattice~\cite{Ejima_2014}. In this process we obtain the central charge $c^*$ very close to the value $c$ obtained in the previous approach (not shown). 

The degeneracy in the entanglement spectrum has been proven to be a suitable measure for probing the topological nature of strongly correlated systems~\cite{Pollmann_2010, Zhou_2023, PhysRevLett.104.130502, Ejima_2014}. In order to properly distinguish the topological (IPM$_\pi$) phase from the trivial (IPM$_0$) phase, we plot four low-lying entanglement eigenvalues ($\varepsilon_1$, $\varepsilon_2$, $\varepsilon_3$ and $\varepsilon_4$) as a function of $J_a$ in Fig.~\ref{fig:entspec}. We observe a two-fold degeneracy of the entanglement spectrum (\textit{i.e.,} $\varepsilon_1=\varepsilon_2$ and $\varepsilon_3=\varepsilon_4$) in the topological BO phase for $J_a>0.5$. On the other hand, no such feature of degeneracy is found in the trivial IPM phase for $J_a<0.5$. \\

\section{Details of the mean-field theory}
\label{appen_mftdetails}

Using the mean-field decoupling defined in Eqs.~\ref{MF_Jb} and \ref{MF_Ja} we can write mean-field form of Hamiltonian in Eq.~\ref{eq_fermionxxz} as 
\begin{widetext}
    \begin{equation}\label{MFT_comb_real_sp_full}
     \begin{split}
         H_{M.F.} =&-\left(\frac{J_\perp}{2} +(1-\alpha)J_a m_e \right)\sum_{I}\left(c_{I,A}^\dag c_{I,B}  +  \text{H.c.} \right) -\left(\frac{J_\perp}{2} +(1-\alpha)J_b m_o \right)\sum_{I}\left( c_{I-1,B}^\dag c_{I,A} +  \text{H.c.} \right) \\
         &+ \alpha(J_a+J_b)\left(n_+ - n_- -\frac{1}{2}\right)\sum_{I} c_{I,A}^\dag c_{I,A}  + \alpha(J_a+J_b)\left(n_+ + n_- -\frac{1}{2}\right)\sum_{I}c_{I,B}^\dag c_{I,B} \\
         & + L\left[\frac{(1-\alpha)}{2}(J_a m_e^2 +J_b m_o^2)+ \alpha(J_a+J_b)(n_-^2 - n_+^2)\right].
     \end{split}
 \end{equation}
\end{widetext}

This then is in momentum space with periodic boundary conditions using the two-site unit cell (we choose the red bonds in Fig.~\ref{fig:fig3} as the unit cell) by defining fermionic annihilation operators for even and odd sites, denoted as  $c_{j,A}$ and $c_{j,B}$ respectively and their Fourier components
\begin{equation}\label{eqn:eq4.7}
       \begin{split}
           c_{j,A/B}=\frac{1}{\sqrt{L}}\sum_{k \in [-\pi/2,\pi/2]} e^{i x_j k}c_{k,A/B},
       \end{split}
\end{equation}
where the summation is over the first BZ corresponding to the two-site unit cell. The Fourier transform of the mean-field Hamiltonian (Eq.~\ref{MFT_comb_real_sp_full}) is 

\begin{widetext}
    \begin{equation}\label{eq:MFT_FT_Comb}
    \begin{split}
        H_{M.F.} =& L\left(\frac{(1-\alpha)}{2}(J_a {m}_e^2+J_b m_o^2)-\alpha (J_a+J_b) \right) n_A n_B\\ &+ \sum_k  \ \ \begin{bmatrix}
c_{k,A}^\dagger & c_{k,B}^\dagger
\end{bmatrix} \ \begin{bmatrix}
 \alpha(J_a+J_b)(n_B  -1/2) & \Delta^c_k\\
\Delta_k^{c*} & \alpha(J_a+J_b)(n_A -1/2)
\end{bmatrix}\begin{bmatrix}
c_{k,A}\\
c_{k,B}
\end{bmatrix},
    \end{split}
\end{equation}
\end{widetext}
where 
\begin{equation}
    \begin{split}
        \label{eq:Dkc}
    \Delta_k^c =& -\frac{J_\perp}{2}(1+e^{2ik})-(1-\alpha)(J_a m_e+J_b m_o e^{2ik}),
    \end{split}
\end{equation}
such that the total ground state energy in terms of $n_+, \ n_-$, $m_o$, and $m_e$ is

\begin{equation}
    \begin{split}
        \label{eq:GS_Energy_full_MFT_comb}
    \frac{E_{G.S.}}{L}=& \frac{(1-\alpha)}{2}(J_a m_e^2+J_b m_o^2)\\
    &+\alpha(J_a+J_b)\left(n_-^2 - n_+^2 + n_+ -\frac{1}{2} \right) \\
    &- \frac{1}{L}\sum_k \sqrt{\alpha^2(J_a+J_b)^2 n_-^2  + |\Delta_k^c|^2}.
    \end{split}
\end{equation}

Minimizing (we can either directly minimize Eq.~\ref{eq:GS_Energy_full_MFT_comb} using \textit{`FindMinimum'} function in Mathematica or can solve it self-consistently, both give exactly the same results) it with respect to $n_+$, $n_-$ $m_o$, and $m_e$ and plotting minimum energy gap (Eq.~\ref{eq:min_gap}) along several cuts for $\alpha=0.6$, we get plot shown in Fig.~\ref{fig:mftpara}(c).
\begin{equation}
    \begin{split}
        \label{eq:min_gap}
    \Delta E (k&=\pi/2) = \\
    &\sqrt{\alpha^2(J_a+J_b)^2 n_-^2  + (1-\alpha)^2 (J_a m_e-J_b m_o)^2}.
    \end{split}
\end{equation}

We determine the order parameter $n_-$ (Figs.~\ref{fig:MFT_PD_comb_nm} and~\ref{fig:MFT_PD_comb_nm_Appendix}) by minimizing the energy (Eq.~\ref{eq:GS_Energy_full_MFT_comb}) for $\alpha=0.5, 0.6, 0.7$. 

\begin{figure}
        \centering

        \includegraphics[width=1.0 \linewidth]{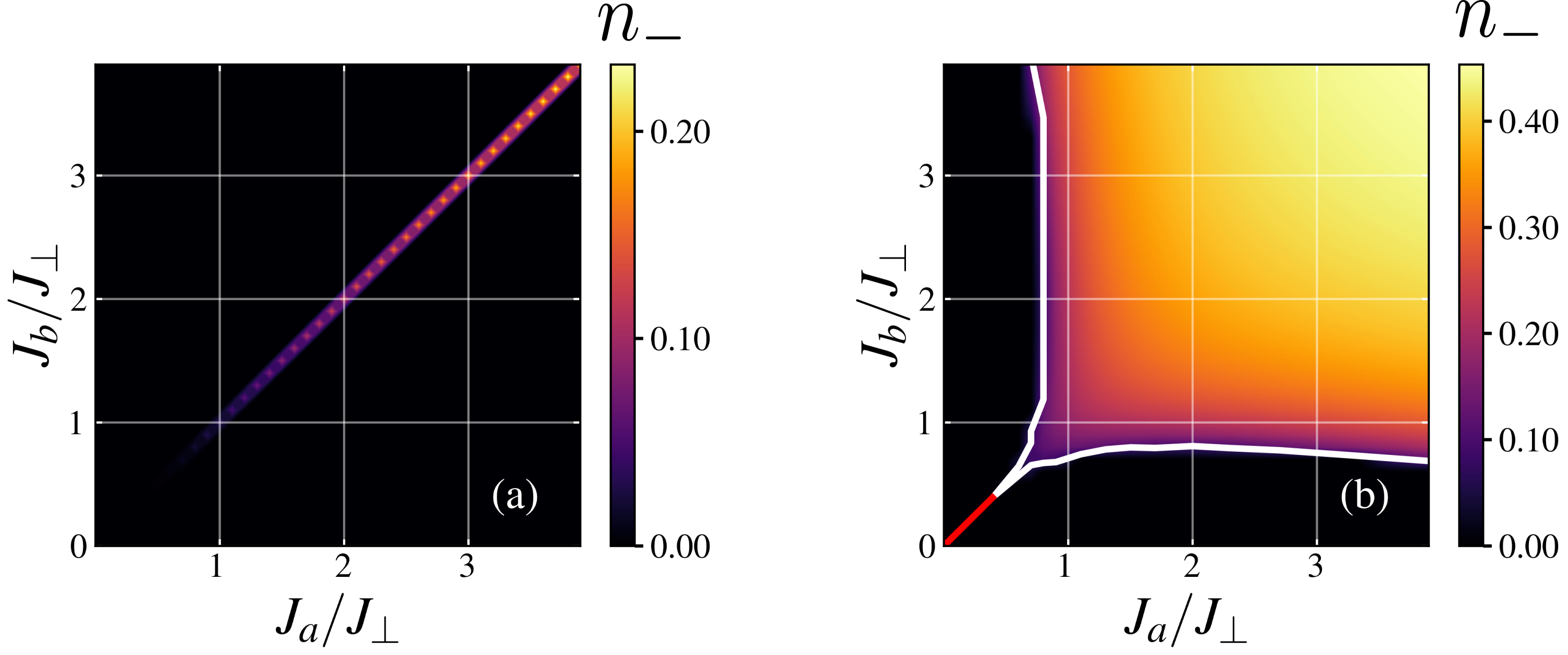}

    \caption{The MF phase diagram obtained via tracking the order parameter $n_-$ and single-particle gap for different values of $(J_a, J_b)$  for $\alpha= 0.5, ~0.7$.}
    \label{fig:MFT_PD_comb_nm_Appendix}
\end{figure}

In the region where $n_-=0$, the MF is reduced to the two bond parameters $m_e$ and $m_o$. In this region, the mean-field Hamiltonian (Eq.~\ref{eq:MFT_FT_Comb}) reduces to (where we have taken $\alpha=0$ below) 

\begin{equation}\label{eqn:eq4.8}
       \begin{split}
           H_{M.F.}(\alpha=0) &=  L \left(\frac{J_b m_o^2+J_a m_e^2}{2}\right)\\
           &+\sum_k \begin{bmatrix}
c_{k,A}^\dagger & c_{k,B}^\dagger
\end{bmatrix} \ \  \ \begin{bmatrix}
 0 & \Delta_k\\
\Delta_k^{*} & 0
\end{bmatrix}   \ \begin{bmatrix}
c_{k,A}\\
c_{k,B}
\end{bmatrix},
       \end{split}
\end{equation}

where 
\begin{equation}\label{eqn:eq4.9}
       \Delta (k)= -\frac{J_\perp}{2}(1+e^{2ik})-(J_a m_e+J_b m_o e^{2ik}),
\end{equation}
which can be solved self-consistently, as before, for $m_o$ and $m_e$. 

\begin{table}
        \centering
        
\begin{tabular}{p{0.10\textwidth} p{0.06\textwidth} p{0.06\textwidth} p{0.06\textwidth} p{0.06\textwidth} p{0.06\textwidth}}
\hline \hline\addlinespace
 Symmetry & $\displaystyle \mathcal{Z}_2^T$ & $\displaystyle U_{z}( 1)$ & $\displaystyle R_{\pi}^{x}$ & $\mathcal{R}_{J_b}$  & $T_{2a}$\\
\hline \addlinespace
 $\displaystyle n_-$ & $\displaystyle -n_-$ & $\displaystyle n_-$ & $\displaystyle -n_-$ & $\displaystyle -n_-$ & $\displaystyle n_- $\\

 $\displaystyle m_o$ & $\displaystyle m_o$ & $\displaystyle m_o$ & $\displaystyle m_o$ & $\displaystyle m_o$ & $\displaystyle m_o$ \\

 $\displaystyle m_e$ & $\displaystyle m_e$ & $\displaystyle m_e$ & $\displaystyle m_e$ & $\displaystyle m_e$ & $\displaystyle m_e$ 
 \\
 \addlinespace \hline \hline
\end{tabular}
        \caption{Symmetry transformations of order parameter $m_0, \ m_e, \ n_-$ under lattice symmetries.}\label{tab:symm_tab_ops}
        \end{table}


\section{\label{appB}Continuum Critical Theory}

To get a continuum theory (in $\lim a\rightarrow 0$) we need to do soft mode expansion of operators $c_i$, $c^\dag_i$ about the two Fermi points, which for half filing is at $k_F=\pm \pi/2$ as shown in Fig.~\ref{fig:low_energy}. We now start from the fermionic Hamiltonian in Eq.~\ref{eq_fermionxxz} which we re-write by using 
\begin{equation}
    J_b = J_0,~~~~~~J_a = J_0 + \delta
\label{eqn:eq2.30c}
\end{equation}
to get
\begin{align}
    H=H_0+H_\delta
\end{align}
where,
\begin{align}\label{eqn:eq2.26}
H_0 =& -\frac{J_\perp}{2}\sum_{i}\left(c_i^\dag c_{i+1} +  \text{H.c.} \right)\nonumber\\
&+J_0\sum_{i}\left(-c_i^\dag c_{i} + c_i^\dag c_{i}c_{i+1}^\dag c_{i+1} \right) 
\end{align}
is the Hamiltonian along the $J_a=J_b$ line and
\begin{align}
    \label{eqn:eq2.32}
    H_{\delta} =  \delta\sum_i c^{\dag}_{2i} c_{2i} c^{\dag}_{2i+1} c_{2i+1}
\end{align}
denotes the dimerization interactions. Now going to momentum space we get:
\begin{align}\label{eqn:eq2.27}
H_0 = &\sum_{k}\xi(k)c_k^\dag c_{k}\nonumber\\
&+ J_0\sum_{k_1,k_2,k_3,k_4} c_{k_1}^\dag c_{k_2}c_{k_3}^\dag c_{k_4} \delta_{k_1 +k_3 -k_2 -k_4,0},
\end{align}
where $\xi(k)=-\left(J_\perp\cos{k}+J_0\right)$. 

Thus, for the free system, when $J_0=0$ system will be at half filling for $k_F=\pi/2$ corresponding to Fig.~\ref{fig:low_energy}. Now we can do a soft-mode expansion of the operators since the low-energy properties are described by continuum field theory of slow varying fields for L/R moving fermions. In continuum limit lattice parameter $a\rightarrow0$, number of sites $L\rightarrow\infty$ such that system length $(aL)$ is finite. In this limit $a\sum_i\rightarrow\int dx$ and 
\begin{equation}\label{eqn:eq2.28}
\frac{\hat{c}_i}{\sqrt{a}}\rightarrow\hat{\psi}(x).
\end{equation}
Expanding $\hat{\psi}(x)$ around $k=\pm k_F$ or L/R points (Fig.~\ref{fig:low_energy}) we get
\begin{equation}\label{eqn:eq2.29}
\hat{\psi}(x)=e^{ik_F x}\psi_R(x)+e^{-ik_F x}\psi_L(x).
\end{equation}

Using Eqs.~\ref{eqn:eq2.28} and~\ref{eqn:eq2.29}, the
Hamiltonian in Eq.~\ref{eqn:eq2.26} can be written \footnote{The continuum limit, $a\rightarrow 0$, cannot be naively taken for 
the quartic interaction terms due to the Pauli exclusion principle.} as (up to an overall energy shift):
\begin{widetext}
    \begin{equation}\label{eqn:eq2.30b_careful}
\begin{split}
    H=&-iv_F\int dx\left[ \psi^\dag_R(x)\partial_x \psi_R(x)- \psi^\dag_L(x)\partial_x \psi_L(x) \right]\\
    &+\lambda\int dx \Big[\left(\rho_R (x) + \rho_L(x)\right)\left(\rho_R (x+a) + \rho_L(x+a)\right) - \psi^\dag_L(x)\psi_R(x) \psi^\dag_R(x+a)\psi_L(x+a) \\
    & ~~~~~~~~~~~~~~~+  \text{H.c.} - \psi^\dag_L(x)\psi_R(x) \psi^\dag_L(x+a)\psi_R(x+a) + \text{H.c.}\Big],
    \end{split}
\end{equation}
\end{widetext}
where $\lambda=\lim_{a\rightarrow 0}(J_0 a)$ denotes the interacting part of the Hamiltonian, $v_F=\lim_{a\rightarrow 0}(J_\perp a)$ is the Fermi velocity and $\rho_r(x) = \psi^\dag_r(x)\psi_r(x)$ is the continuum density for the right and left movers, {\it i.e.} $r=R,L$. The first four terms $\left(\rho_R (x) + \rho_L(x)\right)\left(\rho_R (x+a) + \rho_L(x+a)\right) - \psi^\dag_L(x)\psi_R(x) \psi^\dag_R(x+a)\psi_L(x+a) + \text{H.c.}$  correspond to forward scattering while the last part: $\psi^\dag_L(x)\psi_R(x) \psi^\dag_L(x+a)\psi_R(x+a) + \text{H.c.}$ corresponds to Umklapp scattering.

The soft mode of operators (Eqs.~\ref{eqn:eq2.28} and~\ref{eqn:eq2.29}) in $H_\delta$ now gives,

\begin{widetext}
    \begin{equation}\label{eqn:eq2.33_careful}
\begin{split}
    H_\delta=&\Delta \int dx \Big[\left(\rho_R (x) + \rho_L(x)\right)\left(\rho_R (x+a) + \rho_L(x+a)\right) \\
    &  ~~~~~~~~~~~~- \psi^\dag_L(x)\psi_R(x) \psi^\dag_R(x+a)\psi_L(x+a) + \text{H.c.} 
    - \psi^\dag_L(x)\psi_R(x) \psi^\dag_L(x+a)\psi_R(x+a) + \text{H.c.} \\
    & ~~~~~~~~~~~~- \left(\rho_R (x) + \rho_L(x)\right)\left(\psi^\dag_R(x+a)\psi_L(x+a) + \text{H.c.} \right)\\
    & ~~~~~~~~~~~~+ \left(\rho_R (x+a) + \rho_L(x+a)\right)\left(\psi^\dag_R(x)\psi_L(x) + \text{H.c.}\right)\Big],
    \end{split}
\end{equation}
\end{widetext}
where $\Delta=\lim_{a\rightarrow 0} a\delta/2$. The first eight terms in Eq.~\ref{eqn:eq2.33_careful} are the same as the interaction part in Eq.~\ref{eqn:eq2.30b_careful} so these will modify the coupling strength $\lambda$ in Eq.~\ref{eqn:eq2.30b_careful} to $\lambda+\Delta$. The last eight terms are the new additional terms, which needs regularization as explained in Ref.~\cite{Orignac_Giamarchi} so we  need to normal-order it. After normal ordering and then adding Eqs.~\ref{eqn:eq2.30b_careful} and \ref{eqn:eq2.33_careful}, we get the continuum Hamiltonian for the case $J_a\neq J_b$ as given in Eq.~\ref{eqn:continuum_H_full} in the main text. 

\subsection{\label{App:E}Bosonization}

The bosonic fields $\Phi(x)$ and $\Theta(x)$ introduced in Eq.~\ref{eq:Giamrchi_Bosonization} satisfy the following algebra~\cite{sachdev_2011,Giamarchi_book}
\begin{equation}\label{eqn:eq3.14}
    \begin{split}
        &[\Phi(x),\Theta(y)]=\frac{i\pi}{2}\text{sgn}(x-y),\\
        &[\Phi(x),\nabla\Theta(y)]= i\pi\delta(x-y),\\
        &[\Theta(x),\nabla\Phi(y)]= i\pi\delta(x-y),\\
    \end{split}
\end{equation}
such that $\frac{\nabla\Theta}{\pi}$ is canonically conjugate momenta to $\Phi(x)$, let us denote it by $\Pi_\Phi (x)$. The symmetry transformation of the bosonic fields are given in Table~\ref{tab:bose_field_symm}. The bosonized form of the Hamiltonian (action) is then given by Eq.~\ref{eqn:total_bose_Ht} (\ref{eqn:eq3.32}) in the main text. The bosonized form of the IN and IPM order parameters is given by
\begin{eqnarray}\label{eq:nm_bose}
    &n_{-}\sim \frac{\mathcal{B}}{2}\int dx \langle \cos (2\Phi(x)) \rangle,\\
   \label{Eq:mo_me_bose} &m_{o} \sim -m_{e}\sim -\mathcal{B}\int dx \langle \sin (2\Phi(x)) \rangle.
\end{eqnarray}

\begin{figure}
    \centering
    \includegraphics[width=0.7\columnwidth]{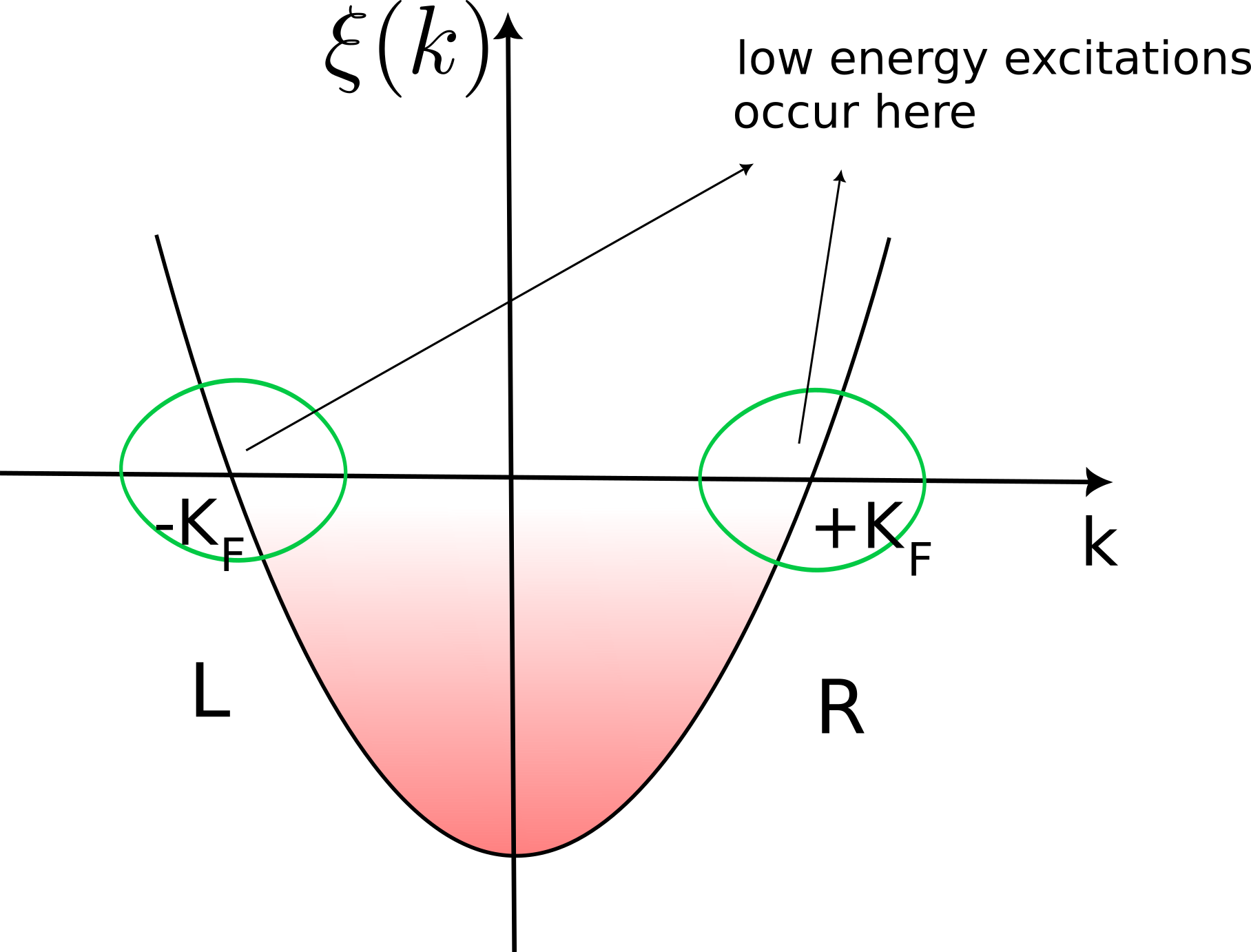}
    \caption{Operators can be expanded in terms of the left (L) and right (R) moving low-energy modes.}
    \label{fig:low_energy}
\end{figure}

To obtain RG flow for couplings $\Delta$ and $\Tilde{\lambda}$, we define the Fourier transform in (1+1)-D Euclidean space as :
\begin{equation}\label{eqn:eq3.33}
   \Phi(x,\tau)=\frac{1}{\beta\Omega}\sum_{k,\omega_n}e^{i(kx-\omega_n\tau)}\Phi(k,\omega_n),
\end{equation}
where $\Omega=L$= spatial volume in 1-D. For simplicity assume $\beta=\infty$ and impose a sharp momentum cutoff $\Lambda$ to start with and then reduce it to  $\Lambda\rightarrow\Lambda/b $ ($b>1$).

Now decompose the field $\Phi(r)$ in fast and slow Fourier modes (\textit{Notation: $r=(x,u\tau)$, $q=(k,\omega_n/u)$}).
\begin{equation}\label{eqn:eq3.34}
   \Phi(r)=\Phi^>(r)+\Phi^<(r),
\end{equation}
where
\begin{equation}\label{eqn:eq3.35}
   \begin{split}
       &\Phi^>(r)=\frac{1}{\beta\Omega}\sum_{\frac{\Lambda}{b}<|q|<\Lambda}e^{iq\cdot r}\Phi^>(q),\\
       &\Phi^<(r)=\frac{1}{\beta\Omega}\sum_{|q|<\frac{\Lambda}{b}}e^{iq\cdot r}\Phi^<(q).
   \end{split}
\end{equation}

\begin{table}

\begin{tabular}{|c|c|}
\hline
Symmetry & Action on fields  \\
\hline\hline
$\mathcal{Z}_2^T$ & $\Phi \mapsto \frac{\pi}{2}-\Phi$, \ \   $\Theta \mapsto \Theta + \pi$  \\
\hline
$U_z(1)$ & $\Phi \mapsto \Phi$, \ \   $\Theta \mapsto \Theta + \chi$  \\
\hline
$R_\pi^x$ & $\Phi \mapsto \frac{\pi}{2}-\Phi$, \ \   $\Theta \mapsto -\Theta$\\
\hline
$\mathcal{R}_{J_b}$ & $\Phi \mapsto \frac{\pi}{2}-\Phi$, \ \   $\Theta \mapsto \Theta + \pi$ \\
\hline
$T_{a}$ & $\Phi \mapsto \Phi + \frac{\pi}{2}$, \ \   $\Theta \mapsto \Theta + \pi$ \\
\hline
\end{tabular}
\caption{Symmetry transformation of the bosonic fields.}\label{tab:bose_field_symm}
\end{table}

To perform the momentum-shell RG, we re-write Eq.~\ref{eqn:eq3.32} as
\begin{align}
    S=S_0+S_p,
\end{align}
where 
\begin{equation}\label{eqn:eq3.37}
   S_0=\frac{1}{2\pi K} \frac{1}{\beta\Omega}\sum_q\left[\frac{\omega_n^2}{u}+uk^2\right]\Phi(q)\Phi^*(q)
\end{equation}

is the quadratic part after taking Fourier transform of it using Eq.~\ref{eqn:eq3.33}, and
\begin{align}
    S_p=\int \frac{d^2r}{u} \mathcal{B}^2 \left[\Delta\sin (2\Phi) - \frac{\tilde{\lambda}}{2} \cos (4\Phi) \right].
    \label{eq_pertaction}
\end{align}

Using $\sum_q = \sum_{\frac{\Lambda}{b}<|q|<\Lambda}+\sum_{|q|<\frac{\Lambda}{b}}$, we get $S_0=S_0^>+S_0^<$ as usual, we can re-write the partition function as
\begin{align}\label{eqn:eq3.38}
   Z&=\int\mathcal{D}[\Phi]e^{-S_0-S_p}\nonumber\\
   &=Z_0^>\int\mathcal{D}[\Phi^<]e^{-S_0^<+ \log \langle e^{-S_p[\Phi^<,\Phi^>]}\rangle_{0>}}\nonumber\\
   &=\int\mathcal{D}[\Phi^<]e^{-S[\Phi^<]},
\end{align}
where the effective action, up to second-order using cumulant expansion, with the reduced cut-off, is given by
\begin{equation}\label{eqn:eq3.40}
   S[\Phi^<]\approx S_0[\Phi^<]+\{\langle S_p\rangle_{0>}-\frac{1}{2}\left(\langle S_p^2\rangle_{0>}-\langle S_p\rangle_{0>}^2\right)\}.
\end{equation}

The first- and second-order corrections are given below. For the 
first-order correction, we have
\begin{align}
    \langle S_p\rangle_{0>}=\langle S_\Delta\rangle_{0>}+\langle S_{\Tilde{\lambda}}\rangle_{0>}
\end{align}
with
\begin{equation}\label{eq:S_Delta}
    S_\Delta = \int \frac{d^2r}{u} \mathcal{B}^2 \Delta \sin (2\Phi),
\end{equation}
and
\begin{equation}\label{eq:S_lambda}
    S_{\Tilde{\lambda}} = -\int \frac{d^2r}{u}  \frac{\mathcal{B}^2}{2} \tilde{\lambda} \cos (4\Phi)
\end{equation}
being the two contributions to $S_p$ in Eq.~\ref{eq_pertaction}. These respectively give
\begin{align}
    \langle S_\Delta\rangle_{0>}=&\mathcal{B}^2 \Delta\int\mathcal{D}[\Phi^>]\frac{e^{-S_0^>}}{Z_0^>} \int\frac{d^2r}{u} \sin{(2\Phi^>(r)+2\Phi^<(r))}\nonumber\\
    =&\mathcal{B}^2 \Delta \int \frac{d^2r}{u}\sin{(2\Phi^<(r)})\ \langle e^{i2\Phi^>(r)}\rangle_{0>} \nonumber\\
    =&\mathcal{B}^2 \Delta  \int\frac{d^2r}{u}\sin{(2\Phi^<(r)})e^{-2\langle \Phi_>^2(r)\rangle_{0>}},
\end{align}
where in the final expression we have used $\langle e^{iA}\rangle=e^{-\langle A^2\rangle/2}$. Similarly, 
\begin{align}
    \langle S_{\Tilde{\lambda}}\rangle_{0>}=-\frac{\mathcal{B}^2}{2}\Tilde{\lambda}\int\frac{d^2r}{u}\cos{(4\Phi^<(r)})e^{-8\langle \Phi_>^2(r)\rangle_{0>}},
\end{align}

where
\begin{equation}\label{eqn:eq3.43}
   \langle \Phi_>^2(r)\rangle_{0>}=\frac{1}{\beta\Omega}\sum_{\frac{\Lambda}{b}<|q|<\Lambda} \frac{\pi K u}{\omega_n^2+u^2k^2}.
\end{equation}

Turning to the second-order correction, we split it into the three contributions as
\begin{align}
    -\frac{1}{2} \Big( \langle S_p^2\rangle_{0>}-\langle S_p\rangle_{0>}^2 \Big)= \mathcal{I}_\Delta + \mathcal{I}_{\Tilde{\lambda}} + \mathcal{I}_{\Delta \Tilde{\lambda}},
\end{align}
where
\begin{align}
    \label{eq:I_d}&\mathcal{I}_\Delta = -\frac{1}{2} \Big( \langle S_\Delta^2\rangle_{0>}-\langle S_\Delta\rangle_{0>}^2 \Big), \\
    \label{eq:I_l}&\mathcal{I}_{\Tilde{\lambda}} = -\frac{1}{2} \Big( \langle S_{\Tilde{\lambda}}^2\rangle_{0>}-\langle S_{\Tilde{\lambda}} \rangle_{0>}^2 \Big), \\
    \label{eq:I_dl}&\mathcal{I}_{\Tilde{\lambda}\Delta} = - \Big( \langle S_{\Tilde{\lambda}}S_\Delta\rangle_{0>}-\langle S_{\Tilde{\lambda}} \rangle_{0>} \langle S_\Delta\rangle_{0>} \Big).
\end{align}

Using Eqs.~\ref{eq:S_Delta},~\ref{eq:S_lambda} and~\ref{eqn:eq3.34} we get:

\begin{widetext}
   \begin{align}\label{eq:S_ld_epsilon}
       & \langle S_\Delta S_{\Tilde{\lambda}} \rangle_{0>} = -\frac{\mathcal{B}^4}{4} \Delta \Tilde{\lambda} \int\frac{d^2r_1}{u}\int\frac{d^2r_2}{u} \sum_{\epsilon = \pm 1} \Big[ \sin{(2\Phi^<(r_2) + \epsilon 4\Phi^<(r_1))}\cdot e^{-\frac{1}{2} \langle  
 [4\Phi^>(r_1) + \epsilon 2\Phi^>(r_2)]^2 \rangle_{0>}}  \Big], \\
  \label{eq:S_d2_epsilon} & \langle S_\Delta^2 \rangle_{0>} = -\frac{\mathcal{B}^4}{2} \Delta^2 \int\frac{d^2r_1}{u}\int\frac{d^2r_2}{u} \sum_{\epsilon = \pm 1} \Big[ \cos{(2\Phi^<(r_1) + \epsilon 2\Phi^<(r_2))}\cdot e^{-2 \langle  
 [\Phi^>(r_1) + \epsilon \Phi^>(r_2)]^2 \rangle_{0>}}  \Big], \\
 \label{eq:S_l2_epsilon} & \langle S_{\Tilde{\lambda}}^2 \rangle_{0>} = \frac{\mathcal{B}^4}{8}  \Tilde{\lambda}^2 \int\frac{d^2r_1}{u}\int\frac{d^2r_2}{u} \sum_{\epsilon = \pm 1} \Big[ \cos{(4\Phi^<(r_1) + \epsilon 4\Phi^<(r_2))}\cdot e^{-8 \langle  
 [\Phi^>(r_1) + \epsilon \Phi^>(r_2)]^2 \rangle_{0>}}  \Big].
   \end{align}
\end{widetext}

Last term in $\mathcal{I}_{\Tilde{\lambda}\Delta}, \mathcal{I}_{\Delta}$ and $\mathcal{I}_{\Delta}$ is there to cancel the disconnected parts (\textit{i.e.} parts for which $r_1$ and $r_2$ are very far from each other). The main contribution thus comes from the region where the two points $r_1$ and $r_2$ are close. There are two main contributions depending upon the sign of $\epsilon$ in Eq.~\ref{eq:S_ld_epsilon} (\ref{eq:S_l2_epsilon}). If $\epsilon=+1$ the term $\sin (2\Phi^<(r_2)+4\Phi^<(r_1))$ ($\cos (4\Phi^<(r_1)+4\Phi^<(r_2))$) in Eq.~\ref{eq:S_ld_epsilon} (\ref{eq:S_l2_epsilon}) will become $\sin (6\Phi^<(r))$ ($\cos (8\Phi^<(r))$) (because we want two points to be close $r_1\sim r_2\sim r$). This term will be a new sine (cosine) term with the argument that is larger than that of the original sine (cosine) and is RG irrelevant and hence can be neglected.  
We can thus only retain contribution with $\epsilon=-1$ in Eqs.~\ref{eq:S_ld_epsilon} and~\ref{eq:S_l2_epsilon}. Now substituting Eq.~\ref{eq:S_ld_epsilon} in Eq.~\ref{eq:I_d} and using the centre of mass coordinate $R=(r_1+r_2)/2$ and relative coordinate $r=r_1-r_2$, we get: 

\begin{widetext}
    \begin{equation}\label{eq:I_dl_comb} 
        \begin{split}
        \mathcal{I}_{\Tilde{\lambda}\Delta} = -\frac{\mathcal{B}^4}{4} \frac{\Delta \Tilde{\lambda}}{u^2}\int d^2R \int d^2r \sin{(2\Phi^<(R))} & \exp{\Big[ -\frac{1}{\beta\Omega} \sum_{\frac{\Lambda}{b}< |q| < \Lambda} \frac{\pi K u}{\omega_n^2 + u^2 k^2}(10-8\cos{(q\cdot r)})   \Big]} \cdot \\
        & \Big( 1 - \exp{\Big[ -\frac{8}{\beta\Omega} \sum_{\frac{\Lambda}{b}< |q| < \Lambda} \frac{\pi K u}{\omega_n^2 + u^2 k^2} \cos{(q\cdot r)}  \Big]}\Big). 
        \end{split}
    \end{equation}
\end{widetext}

We calculate the summations inside the exponentials in the Eq.~\ref{eq:I_dl_comb} by approximating these by an integral and using a soft-cutoff procedure ($\int_0^\Lambda d^2q \rightarrow \int_0^\infty d^2q \Lambda^2/(q^2 + \Lambda^2)$). Finally, doing rescaling ($r=br'$ or $q=q'/b$ and $b=e^l$) and renormalization ($\Phi^<(r')\rightarrow \Phi'(r')$)  we get:

\begin{equation}\label{eq:I_dl_final}
    \mathcal{I}_{\Tilde{\lambda}\Delta} = - c_1  \Delta \Tilde{\lambda} \delta l \mathcal{B}^2 \int d^2r' \sin{2\Phi'(r')}.
\end{equation}

Using Eqs.~\ref{eq:S_Delta}, \ref{eq:S_lambda} and \ref{eq:I_dl_final} in Eq.~\ref{eqn:eq3.40} leads to the 
$\beta$-function (up to second order) for the coupling $\Delta$,
\begin{equation}
    \frac{d\Delta}{dl} = (2-K)\Delta - c_1 \Delta \Tilde{\lambda},
\end{equation}
where 
\begin{equation}
    c_1 = \frac{4K}{\pi u (\Lambda a)^2}.
\end{equation}

$\mathcal{I}_{\Delta}$ can be evaluated in a similar manner and it will lead to $\beta$-function expression for $\Tilde{\lambda}$:

\begin{equation}
    \frac{d\Tilde{\lambda}}{dl} = (2-4K)\Tilde{\lambda} - c_1 \Delta^2.
\end{equation}

Similarly, we get the $\beta$-function for the Luttinger parameter as 
\begin{equation}\label{eq:K-beta}
    \frac{dK}{dl} = -a_1(\Delta^2 + 4\Tilde{\lambda}^2),
\end{equation}
where,
\begin{equation}
    a_1 = \frac{32 K^3}{(\pi u (\Lambda a)^2)^2}.
\end{equation}
These can be cast in the final form given by Eq.~\ref{eqn:total_betat}.
\bibliography{references}

\end{document}